\documentclass[aps,prl,reprint,superscriptaddress,showpacs,longbibliography,footnoteinbib]{revtex4-1}
\usepackage[utf8]{inputenc}
\usepackage{color}
\usepackage{bm}
\usepackage{amsmath}
\usepackage{amssymb}
\usepackage{graphicx}
\usepackage[unicode=true,
 bookmarks=false,
 breaklinks=false,pdfborder={0 0 1},backref=false,colorlinks=true]
 {hyperref}
\hypersetup{pdftitle={Title},
 plainpages=false,pdfpagelabels,linkcolor=blue,urlcolor=blue,citecolor=blue,pdfdisplaydoctitle=true,pdfduplex=DuplexFlipLongEdge}

\makeatletter

\providecommand{\tabularnewline}{\\}

\usepackage{units}\usepackage{wasysym}

\newcommand{\beginsupplement}{%
	\setcounter{page}{1}
	 \renewcommand{\thepage}{SM - \arabic{page}}%
        \setcounter{table}{0}
        \renewcommand{\thetable}{S\arabic{table}}%
        \setcounter{figure}{0}
        \renewcommand{\thefigure}{S\arabic{figure}}%
        \setcounter{section}{0}
        \renewcommand{\thesection}{S\arabic{section}}%
        \setcounter{section}{0}
        \renewcommand{\thesection}{S\arabic{section}}%
        \setcounter{subsection}{0}
        \renewcommand{\thesubsection}{S\arabic{section}.\arabic{subsection}}%
        \setcounter{equation}{0}
        \renewcommand{\theequation}{S\arabic{equation}}%

     }

\usepackage{bm, soul}
\usepackage{braket}

\makeatother

\begin{document}
\noindent
\title{%Non-interacting universality of 
Short-range interactions are irrelevant at
the quasiperiodic-driven Luttinger Liquid to Anderson Glass transition}
%\title{Non-interacting universality of 
%
%the quasiperiodic-driven Luttinger Liquid to Anderson Glass transition}

\author{Miguel Gonçalves}
\affiliation{CeFEMA, Instituto Superior Técnico, Universidade de Lisboa, Av. Rovisco
Pais, 1049-001 Lisboa, Portugal}
\author{J. H. Pixley}
\affiliation{Department of Physics and Astronomy, Center for Materials Theory,
Rutgers University, Piscataway, New Jersey 08854, USA}
\affiliation{Center for Computational Quantum Physics, Flatiron Institute, 162
5th Avenue, New York, New York 10010, USA}
\author{B. Amorim}
\affiliation{Centro de Física das Universidades do Minho e do Porto (CF-UM-UP), 
Laboratório de Física para Materiais e Tecnologias Emergentes (LaPMET), Universidade do Minho, 4710-057 Braga, Portugal}
\author{Eduardo V. Castro}
\affiliation{Centro de Física das Universidades do Minho e Porto, Departamento
de Física e Astronomia, Faculdade de Ciências, Universidade do Porto,
4169-007 Porto, Portugal}
\affiliation{Beijing Computational Science Research Center, Beijing 100193, China}
\author{Pedro Ribeiro}
\affiliation{CeFEMA, Instituto Superior Técnico, Universidade de Lisboa, Av. Rovisco
Pais, 1049-001 Lisboa, Portugal}
\affiliation{Beijing Computational Science Research Center, Beijing 100193, China}
\begin{abstract}
We show that short-range interactions are irrelevant around gapless ground-state
delocalization-localization transitions driven by quasiperiodicity
in interacting fermionic chains. In the presence of interactions,
these transitions separate Luttinger Liquid and Anderson glass phases. Remarkably,
close to criticality, we find that excitations become effectively
non-interacting. By formulating a many-body generalization of a recently developed method to obtain single-particle
localization phase diagrams, we carry out precise calculations of critical points between Luttinger Liquid and Anderson glass phases 
and find that the correlation length critical exponent takes the value $\nu = 1.001 \pm 0.007$, compatible with $\nu=1$ known exactly 
at the non-interacting critical point. We also show that other critical
exponents, such as the dynamical exponent $z$ and a many-body analog
of the fractal dimension are compatible with the exponents obtained at the non-interacting critical point.
Noteworthy, we find that the transitions are accompanied by the emergence of a
many-body generalization of previously found single-particle hidden dualities. Finally, we show that in the limit of vanishing interaction strength, all finite range interactions are irrelevant at the non-interacting critical point.

\end{abstract}
\maketitle

There has been a continued interest in the effects of quasiperiodicity on quantum many body systems thanks to their experimental accessibility in ultracold atoms and  trapped ions \citep{PhysRevA.75.063404,Roati2008,Modugno_2009,Schreiber842,Luschen2018,PhysRevLett.123.070405,PhysRevLett.125.060401,PhysRevLett.126.110401,PhysRevLett.126.040603,PhysRevLett.122.170403} and most recently moir\'e materials  \citep{Balents2020}. From a theoretical point of view, the effects of interactions on  Anderson insulating ground states is also of paramount importance in the context of many-body localization, where random and quasiperiodic systems have important and fundamental differences that are currently under intense scrutiny \citep{PhysRevLett.119.075702,PhysRevB.96.104205,PhysRevX.7.031061,Agrawal2020}. Of paramount importance is understanding the nature of such many body localization phase transitions that take place at finite energy density far away from the ground state \citep{PhysRevB.87.134202,PhysRevA.92.041601,PhysRevLett.115.230401,PhysRevB.96.075146,Znidari,PhysRevResearch.1.032039,PhysRevB.100.104203,PhysRevLett.128.146601,PhysRevB.104.214201}. However, even in the limit of the ground state, the nature of the universality class of the interacting quasiperiodic electronic ``glass'' transition   has remained poorly understood.
A great deal of understanding has been achieved in  non-interacting
quasiperiodic systems thanks to rigorous results on the paradigmatic Aubry-André model \citep{AubryAndre,Avila10Martini, 
 PhysRevB.98.134201,PhysRevB.101.174203},
where an energy-independent delocalization-localization takes place,
and on its generalizations to phase diagrams that contain mobility edges
and/or critical phases \citep{PhysRevB.43.13468,PhysRevLett.104.070601,PhysRevLett.113.236403,Liu2015,PhysRevB.91.235134,PhysRevLett.114.146601,anomScipost}.
However, the interplay between quasiperiodicity and interactions has
been less explored. Typically, the studies on this interplay are in the context
of many-body localization, for highly excited states in the middle
of the many-body spectrum \citep{PhysRevB.87.134202,PhysRevA.92.041601,PhysRevLett.115.230401,PhysRevB.96.075146,Znidari,PhysRevResearch.1.032039,PhysRevB.100.104203,PhysRevLett.128.146601,PhysRevB.104.214201}.
An equally interesting direction is the study of ground-state localization
properties 
\citep{PhysRevLett.83.3908,PhysRevB.65.115114,PhysRevA.78.023628,PhysRevB.89.161106,SciPostPhys.1.1.010,PhysRevB.101.174203,PhysRevLett.126.036803,PhysRevX.7.031061,PhysRevLett.120.175702,Agrawal2020,Crowley_2022,PhysRevLett.126.036803,PhysRevB.106.L121103}.
Deep enough in the localized phase, upon adding nearest-neighbor interactions to  Aubry-André model, the ground-state remains localized, giving
rise to an Anderson glass (AG) phase \citep{PhysRevB.65.115114,SciPostPhys.1.1.010,Mastropietro2017,PhysRevB.101.174203}.
On the other hand, at weak interaction strength the Luttinger liquid (LL) phase  is stable towards the inclusion of a sufficiently weak quasiperiodic potential
\citep{PhysRevB.65.014201,SciPostPhys.1.1.010}. The gapless ground-state
delocalization-localization transition  therefore persists
in the presence of interactions, corresponding to a transition between
the LL and AG phases. The critical properties of the LL-AG transition were studied
in detail for the non-interacting Aubry-André model \citep{PhysRevB.98.134201,PhysRevA.99.042117,PhysRevB.101.174203},
and it was proposed in \citep{PhysRevB.101.174203} that nearest-neighbor
repulsive interactions may be irrelevant at the critical point. In
more generic interacting models beyond the paradigmatic Aubry-André model, the ground-state
localization properties remain largely unexplored. 

%\jp{JP: Too many acronyms} \miguel{MG: Only left LL and AG acronyms}

\begin{figure*}
\centering{}\includegraphics[width=1\textwidth]{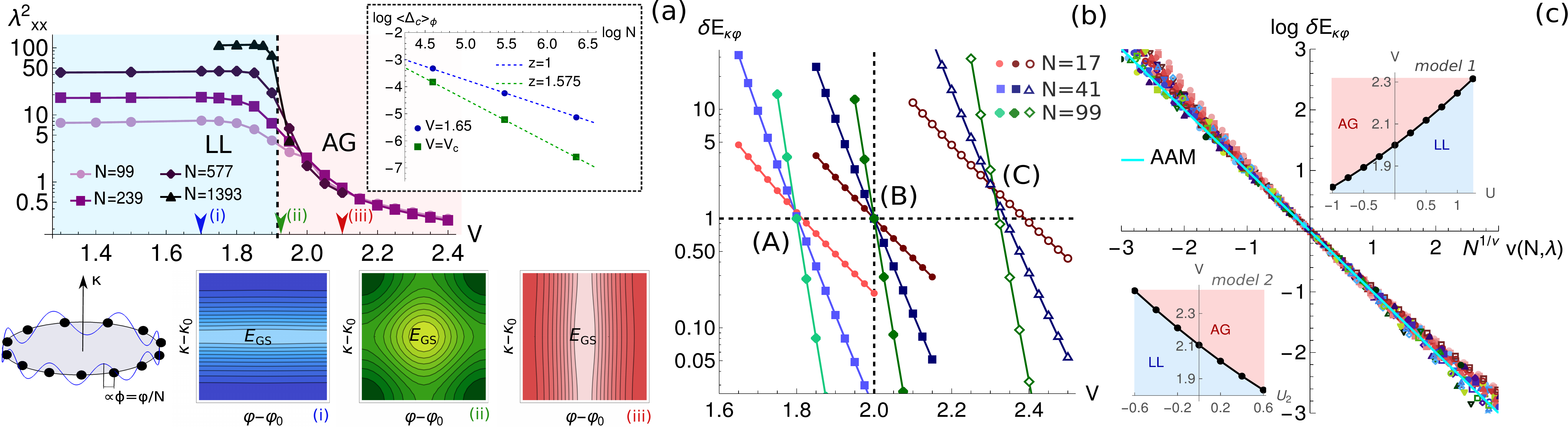}\caption{{\bf Quantum phase transition between Luttinger liquid (LL) and Anderson glass (AG) phases.} (a) The many-body localization tensor $\lambda_{xx}^{2}$, as given in Eq.$\,$\ref{eq:Kohn} and computed using open boundary
conditions for different system sizes $N$, is plotted versus the quasiperiodic potential strength $V$, for
the model in Eq.$\,$\ref{eq:H} with $V_{2}=0.25,t_{2}=0.2,U=0.5$
and $U_{2}=0.4$. 
The scaling of the $\phi$-averaged charge gap  with system size  (see Eq.$\,$\ref{cgap} and Fig.$\,$\ref{fig:3}(c) for definition and additional details) in the LL phase (for $V=1.65$) and at the critical point (indicated by the vertical dashed line) is shown in the inset, unveiling a non-Lorentz-invariant critical point (with dynamical critical exponent $z > 1$). The blue
and green dashed lines show the scaling behaviour known for the non-interacting Aubry-André model, respectively at extended and
critical points, which is compatible with the scaling behaviour observed at the LL phase and the interacting critical point.
Below these figures, we show
the $E_\text{GS}(\varphi,\kappa)$ contours for a periodic system with a threaded flux $\kappa$ (illustrated in the bottom left figure), for $N=41$, at representative points in the LL (i) and AG (iii) phases, and at
the self-dual point (ii), that approaches the critical point (whose estimation
is shown by the vertical dashed line) as $N\rightarrow\infty$. $(\varphi_{0},\kappa_{0})$
in these figures is chosen so that $E_\text{GS}(\varphi_{0},\kappa_{0})$ is
minimum. (b) $\delta E_{\kappa\varphi}$ defined in Eq.$\,$\ref{eq:DeltaEkphi}
(using $\delta s=\pi/20$) for (A): model 1 (see main text bellow Eq.$\,$\ref{eq:H}), with $U=-1$; (B) Aubry-André model
(C) model 2 with $U_{2}=-0.4$, for $L=\{17\textrm{(red)},41\textrm{(blue)},99\textrm{(green)}\}$.
(c) Data collapse of $\log\delta E_{\kappa\varphi}$ as a function
of $N^{1/\nu}v(N,\bm{\lambda})$, using $\nu=1$, where $\bm{\lambda}$
contains the model parameters, $v(N,\bm{\lambda})=[V-V_{c}(N,\bm{\lambda})]/V_{c}(N,\bm{\lambda})$
and $V_{c}(N,\bm{\lambda})$ is the value of $V$ for which $\delta E_{\kappa\varphi}(N,V,\bm{\lambda})=1$.
The data was obtained across the LL-AG transitions shown in the insets,
for model 1 (right) and model 2 (left). The results for $N=\{17,41,99\}$
correspond to the red, blue and green markers, respectively. In cyan,
we show the result for the non-interacting Aubry-André model. The critical points
in the insets {[}as well as the critical point in (a){]} are estimated
through $V=V_{c}(N=99,\bm{\lambda})\approx V_{c}(\infty,\bm{\lambda})$.
The validity of this estimation is confirmed by the small error bars
(smaller or of the size of the data points), computed through $|V_{c}(99,\bm{\lambda})-V_{c}(41,\bm{\lambda})|$.\label{fig:1}}
\end{figure*}

In this paper we show that interactions are irrelevant around quasiperiodicity-driven
LL-AG transitions for a broad class of non-interacting and interacting generalizations
of the Aubry-André model, that include next-nearest-neighbor hoppings and interactions
and an additional quasiperiodic potential. In particular, we show
that the excitations become effectively non-interacting around these
transitions and provide solid evidence that the addition of interactions
does not affect any of the (infinite number of) critical exponents obtained in the non-interacting
limit, although they modify non-universal properties, e.g.  the location of the critical point. Remarkably, we
also find that many-body generalizations of the single-particle dualities
discovered for widely different 1D models \citep{HdualitiesScipost,PhysRevLett.104.070601,PhysRevLett.114.146601}
emerge around criticality. 
We present a scaling argument based on the perturbative effects of interactions at the Aubry-Andre critical point which demonstrates that short-range interactions are irrelevant. The same argument shows that long-range interactions can become relevant. 
%In the limit of vanishing interaction strength, we show that more generic short-range interactions are irrelevant at the non-interacting critical point.

Our main results are shown in Fig.$\,$\ref{fig:1}. In Fig.$\,$\ref{fig:1}(a)
we show an example of a LL-AG transition, in which
we set all the couplings of our class of models (in Eq.$\,$\ref{eq:H}
below) different from zero. A way to capture this transition is with
the many-body localization length (see Eq.$\,$\ref{eq:Kohn} for definition) \citep{PhysRevLett.82.370,Resta2011,PhysRevB.92.134207} that diverges (saturates) at the LL (AG) phase due to the extended (localized) nature of the many-body wave function. 
 
One of our main findings in this work is that
this transition can also be captured in a  precise manner with minimal scaling assumptions, using a many-body generalization of the single-particle
theory developed in \citep{RG_paper,10.21468/SciPostPhys.13.3.046}.
This involves considering periodic approximations of the quasiperiodic
system, as illustrated at the bottom left corner of Fig.$\,$\ref{fig:1}(a),
and inserting a flux $\kappa$ through the resulting ring. The
localization properties can then be inferred based on how the ground-state
energy ($E_{{\rm GS}}$) depends on these fluxes and on real-space
shifts between the lattice %sites 
and origin of the potential that we encode
in the variable $\varphi$, as illustrated in Fig.$\,$\ref{fig:1}(a),
as the size of the periodic approximation ($N$) is increased. The
quasiperiodic limit is approached for $N\rightarrow\infty$.  In Fig.$\,$\ref{fig:1}(a), we show examples of the energy contours for fixed $N=41$, for different values of $V$. We can see that in the LL (AG) phase, there is a very small dependence on $\varphi$ ($\kappa$), while close to the critical point, there is an equal dependence on both phases.  
We can make a more quantitative analysis by computing the ratio between the energy dispersions along $\kappa$
and $\varphi$ respectively, $\delta E_{\kappa\varphi}$ (see Eq.$\,$\ref{eq:DeltaEkphi}
for precise definition), for different $N$. Example results are given in Fig.$\,$\ref{fig:1}(b) for different models, where we can see that $\delta E_{\kappa\varphi}$ diverges (scales to zero) at the LL (AG)
phase. This implies that the
$\varphi$-($\kappa$-) dependence becomes irrelevant with respect
to the $\kappa$-($\varphi$-) dependence at the LL (AG) phases as $N$ increases, and the model flows to a delocalized (localized) fixed-point as defined in \citep{GoncalvesRG2022}.
Remarkably, at the critical point, $\delta E_{\kappa\varphi}$ approaches unity as $N$ is increased, which implies that the shift and flux dependencies become equivalent. The \textit{self-dual point}, defined here as the point for which $\delta E_{\kappa\varphi}=1$ \footnote{This matches the usual definition of a self-dual point, when an exact duality transformations can be explicitly constructed.}, therefore approaches the critical point as $N\rightarrow \infty$, providing a very precise way of estimating it.

With the ansatz $\delta E_{\kappa\varphi}\propto e^{-N/\xi}$,
where $\xi\sim |V-V_c|^{-\nu}$ is the correlation length, we were able to collapse
the results 
%for widely different models and different system sizes
for each model considered here
into a single universal curve using the correlation length critical
exponent $\nu=1$ known for the non-interacting Aubry-André model, as shown in Fig.$\,$\ref{fig:1}(c).
The obtained scaling function near criticality is 
in excellent agreement with
%to 
the
one obtained for the non-interacting Aubry-André model, shown in cyan in Fig.$\,$\ref{fig:1}(c).
The good quality of the collapse and further results on additional critical exponents
presented throughout the manuscript support the conclusion that around
criticality, different interacting and non-interacting models belong
to the same non-interacting universality class.

\begin{figure}[h]
\centering{}\includegraphics[width=1\columnwidth]{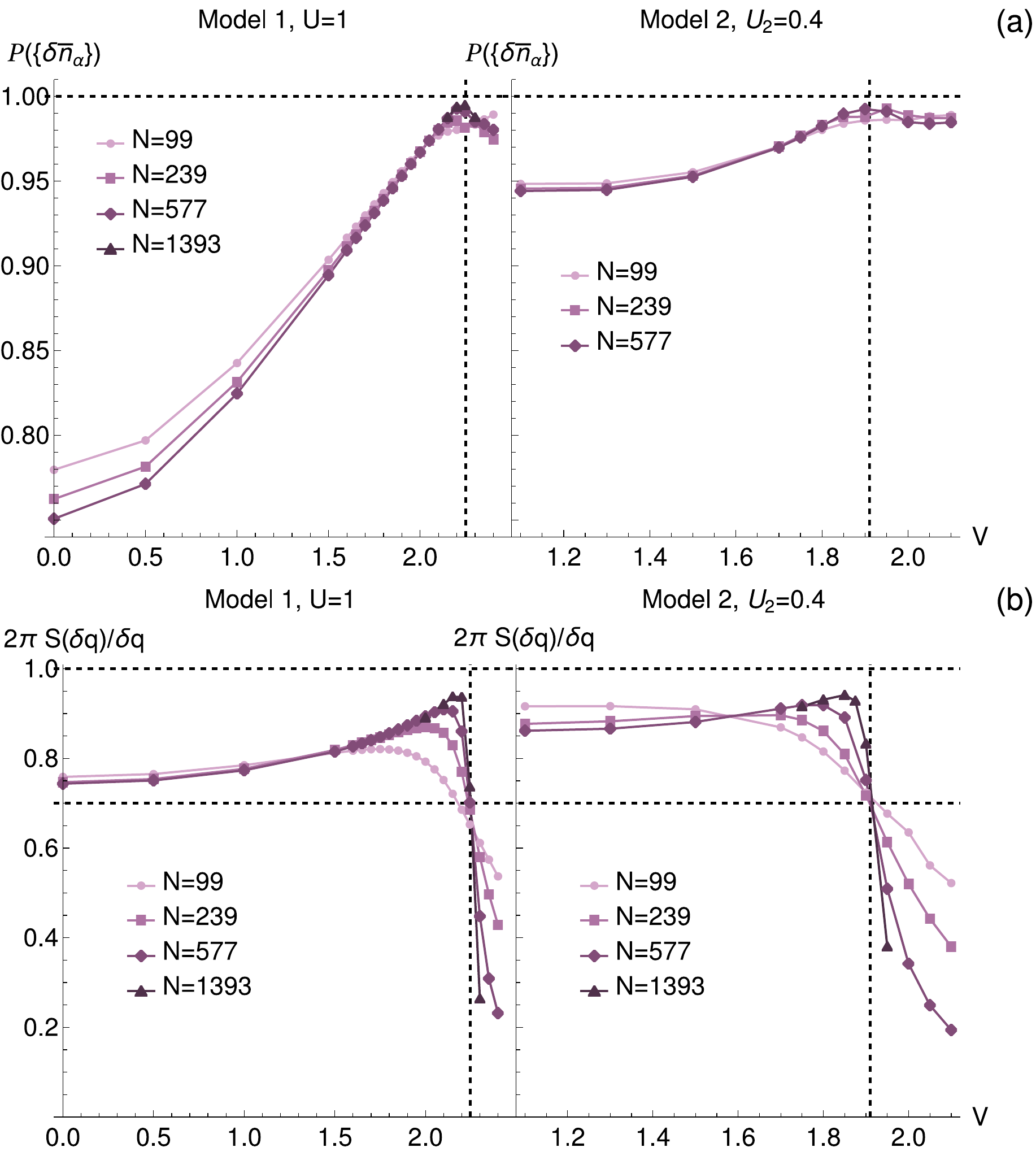}\caption{
\textbf{Non-interacting excitations at the critical point.}
(a) Occupation inverse participation ratio $P(\{\delta\bar{n}_{\alpha}\})$ defined below Eq.$\,$\ref{excitation_matrix}, and (b) slope of structure factor $S(\delta q)$ for the smallest non-vanishing momentum $\delta q = 20\pi/N$ as a function of $V$ for $U=1,\phi=1.123$ and
$V_{2}=t_{2}=U_{2}=0$ (left) and $U=0.5,V_{2}=0.25,t_{2}=0.2,U_{2}=0.4$
(right). The vertical dashed line corresponds to the critical point
estimated from $\delta E_{\kappa\varphi}$ and the horizontal dashed
line is the slope of the linear contribution to $S(q)$ at the critical point (see text).
\label{fig:2}}
\end{figure}

\paragraph*{Models and Methods.--- }

We study the class of models described by the Hamiltonian 
\begin{equation}
\begin{aligned}H= & -t\sum_{i}c_{i}^{\dagger}c_{i+1}-t_{2}\sum_{i}c_{i}^{\dagger}c_{i+2}+{\rm h.c.}\\
 & +\sum_{i}\Big(V\cos(2\pi\tau i+\phi)+V_{2}\cos[2(2\pi\tau i+\phi)]\Big)c_{i}^{\dagger}c_{i}\\
 & +U\sum_{i}n_{i}n_{i+1}+U_{2}\sum_{i}n_{i}n_{i+2}
\end{aligned}
\label{eq:H}
\end{equation}
where $c_{i}^{\dagger}$ creates a particle at site $i$ and we set
$t=1$ throughout the manuscript. The first and third rows contain
nearest- and next-nearest-neighbor hoppings and interactions, respectively,
while the second row contains quasiperiodic potentials of intensity
$V$ and $V_{2}$, with $\tau$ being an irrational number and the
phase $\phi$ representing a shift of the potentials with respect
to the lattice sites. For the results presented in this manuscript,
we set $\tau=1/\sqrt{2}$ (to compare results with Refs.$\,$\citep{PhysRevB.101.174203,SciPostPhys.1.1.010})
and work at half-filling $\rho=1/2$ (unless otherwise specified), choosing a number of particles
$N_{p}=\lfloor N/2\rfloor$, where $\lfloor x\rfloor$ denotes the
integer part of $x$ (the floor function). 
We have checked that our  conclusions do not rely on being at this particular filling, see SM.
We choose the following sets of parameters:
Aubry-André model with nearest-neighbor interaction, $V_{2}=t_{2}=U_{2}=0$ (\textit{model
1}) ; generalized Aubry-André model, with $U=0.5,V_{2}=0.25,t_{2}=0.2,U_{2}\neq0$
(\textit{model 2}).

To study the models in detail, we computed several different quantities
using the DMRG technique \citep{PhysRevLett.69.2863,RevModPhys.77.259},
as implemented in the iTensor library \citep{itensor,itensorpaper},
applying both periodic and open boundary conditions. 
We required iTensor's truncation error to be less than $10^{-10}$ and only stopped the sweeping procedure once some convergence requirements were satisfied, up to a maximum of $500$ sweeps. In particular, for twisted boundary conditions, we require the energy variance, $\Delta_{H}=\langle H^{2}\rangle-\langle H\rangle^{2}$, to be below $5 \times 10^{-6}$; the ground-state energy difference between two sweeps to be below $\Delta E_{{\rm GS}}=10^{-5}$; and the difference in the entanglement entropy at the middle bond inbetween two sweeps to be below $\Delta S_{{\rm GS}}=10^{-3} $. For open boundary conditions, we require $\Delta_{H}$ at least below $10^{-6}$; $\Delta E_{{\rm GS}}\leq 10^{-7}$; and $\Delta S_{{\rm GS}} \leq 10^{-4} $.

In our finite-size simulations, we use rational approximants of $\tau$,
$\tau_{c}=p/N$, with $p$ and $N$ co-prime numbers. These approximants
were chosen to be exact convergents of the continued fraction expansion
of $\tau$. 

\paragraph*{\textmd{Twisted boundary conditions.---} }

We consider a ring with $N$ sites as illustrated in the bottom left
corner of Fig.$\,$\ref{fig:1}(a) with twisted boundary conditions that corresponds to  threading a flux $\kappa$ through the system.
This flux can simply be added to the model in Eq.$\,$\ref{eq:H}
by making the replacement $t\rightarrow te^{ik}$ and $t_{2}\rightarrow t_{2}e^{2ik}$,
with $k=\kappa/N$. For the choices $\tau=\tau_{c}$, making shifts
$\phi\rightarrow\phi+2\pi/N$ simply corresponds to a relabelling
of the indices in this model \citep{10.21468/SciPostPhys.13.3.046},
which implies that the many-body ground-state energy is periodic in
$\phi$ with period $\Delta\phi=2\pi/N$. With this in mind, we define
the rescaled variable $\varphi=N\phi$ so that $E_{{\rm GS}}(\varphi,\kappa)$
has a period $\Delta\varphi=2\pi$. We define the flux-shift sensitivity,
$\delta E_{\kappa\varphi}$, as 
\begin{equation}
\delta E_{\kappa\varphi}=\lim_{\delta s\rightarrow0}\frac{E_{{\rm GS}}(\varphi_{0},\kappa_{0}+\delta s)-E_{{\rm GS}}(\varphi_{0},\kappa_{0})}{E_{{\rm GS}}(\varphi_{0}+\delta s,\kappa_{0})-E_{{\rm GS}}(\varphi_{0},\kappa_{0})}\label{eq:DeltaEkphi}
\end{equation}
where $(\varphi_{0},\kappa_{0})$ are defined so that $E(\varphi_{0},\kappa_{0})$
is minimum to ensure that $\delta E_{\kappa\varphi}=1$ at self-dual points
\footnote{Note that $(\varphi_{0},\kappa_{0})$ should be chosen so that at
self-dual points, the energy dispersions are invariant under switching $\varphi-\varphi_{0}$
and $\kappa-\kappa_{0}$.}. Note that the values of $(\varphi_{0},\kappa_{0})$ can depend on
$N$ and on the number of particles $N_{p}$. For the system sizes
used in the calculations with periodic boundary conditions and $N_{p}=\lfloor N/2\rfloor$,
we found $(\varphi_{0},\kappa_{0})=(0,\pi)$ for $N=17,41$ and $(\varphi_{0},\kappa_{0})=(\pi,0)$
for $N=99$. $\delta E_{\kappa\varphi}$ is the many-body generalization
of a similar quantity already introduced for the single-particle eigenenergies
of non-interacting quasiperiodic models in \citep{10.21468/SciPostPhys.13.3.046}.
In the LL (AG) phase, we expect $\delta E_{\kappa\varphi}\rightarrow\infty$
($\delta E_{\kappa\varphi}\rightarrow0$) for increasing $N$. At
the critical point, we have $\delta E_{\kappa\varphi}\rightarrow1$,
as we shall see.

It is clear from the $E_{{\rm GS}}(\varphi,\kappa)$ contour plots
in Fig.$\,$\ref{fig:1} that there is a duality between the LL and
AG phases around criticality under switching $(\varphi-\varphi_{0})$
and $(\kappa-\kappa_{0})$. The critical point is the self-dual point
of this duality, in which $E_{{\rm GS}}$ is invariant under this
exchange. In Ref.$\,$\citep{10.21468/SciPostPhys.13.3.046} we have
uncovered similar dualities in the single-particle case and found
that they could be traced-back to hidden duality transformations between
the single-particle wave functions. Remarkably, in the presence of interactions, a many-body generalization of these duality transformations can still be formulated. In the SM we provide the precise definition and some examples.

\paragraph*{\textmd{Open boundary conditions.---} }

We also employ open boundary conditions that allow to reach fairly
large system sizes \citep{itensor,itensorpaper}. In order to carry
out a complete study of the LL and AG phases, and of the transition between them, we
compute several quantities that we detail below.  

The many-body localization length shown in Fig.$\,$\ref{fig:1} can be defined as \citep{PhysRevLett.82.370,Resta2011,PhysRevB.92.134207}
\begin{equation}
\lambda_{xx}^{2}=(\bra{\Psi}\hat{x}^{2}\ket{\Psi}-\bra{\Psi}\hat{x}\ket{\Psi}^{2})/N_{p},\label{eq:Kohn}
\end{equation}
where $\hat{x}=\sum_{i=1}^{N_{p}}\hat{x}_{i}=\sum_{i=1}^{N}x_{i}c_{i}^{\dagger}c_{i}$
is the many-body position operator and $N_{p}$ is the number of particles.  Since it measures the variance of the position operator, it can distinguish between the LL and AG phases: it diverges (saturates) with $N$, due to the extended (localized) nature of the many-body wave function at the LL (AG) phase.

To verify the gapless nature of the transition and obtain the dynamical critical exponent $z$, we also computed the charge gap, defined as 
\begin{equation}
\Delta_{c}=E_{{\rm GS}}(N_{p}+1)+E_{{\rm GS}}(N_{p}-1)-2E_{{\rm GS}}(N_{p}) \sim N^{-z},
\label{cgap}
\end{equation}

\noindent where $E_{{\rm GS}}(N_{p})$ is the ground-state energy for $N_{p}$
particles.

In order to study the scaling of the entanglement in different regions of the phase diagram, we computed the entanglement entropy \citep{PhysRevLett.90.227902,RevModPhys.80.517},
defined as 
\begin{equation}
\mathcal{S}=-\rm{Tr}[\rho_{A}\log\rho_{A}],\textrm{ }\rho_{A}=\rm{Tr}_{B}\ket{\psi}\bra{\psi},
\end{equation}
where we choose the partition $A$
containing the first $N_{A}$ sites of the chain. For a 1D critical
system whose continuum limit is a conformal field theory with central
charge $c$, we have that \citep{Calabrese_2004}
\begin{equation}
\mathcal{S}=\frac{c}{6}\log\Big(\frac{N}{\pi}\sin(\pi N_{A}/N)\Big)+C'. \label{eq:S_calebrese}
\end{equation}
This is the expected behaviour at the LL phase (with $c=1$), while at the AG phase, $\mathcal{S}$ becomes non-extensive for large enough $N_A$.

To better understand the nature of single-particle
excitations we also introduce here the \textit{particle-addition correlation matrix}, that we define as 
\begin{equation}
C_{e}^{ij}=\langle c_{i}^{\dagger}c_{j}\rangle_{N_{p}}-\langle c_{i}^{\dagger}c_{j}\rangle_{N_{p}-1}
\label{excitation_matrix}
\end{equation}
where $\langle\rangle_{N_{p}}$ denotes expectation value in the ground-state
with $N_{p}$ particles. The eigenvalues and eigenvectors of the particle-addition correlation 
matrix, $\bm{C}_{e}\ket{\alpha}=\delta\bar{n}_{\alpha}\ket{\alpha}$, with $\alpha=0,\cdots,N-1$, 
correspond to the occupations and natural orbitals. For a non-interacting system,
only a single natural orbital corresponding to the $N_{p}$-th highest-energy single-particle eigenstate labelled as $\ket{\alpha=0}$ is occupied (we have $\delta\bar{n}_{0}=1;\delta\bar{n}_{\alpha>0}=0$). In contrast, in the presence of interactions, a particle that is added to the system redistributes over different natural orbitals. Therefore, the deviations from the expected behaviour of a non-interacting particle can be quantified by inspecting the occupations $\{ \delta\bar{n}_{\alpha} \}$. For this purpose, we  
introduce the \textit{occupation inverse participation ratio} defined as $P(\{\delta\bar{n}_{\alpha}\})=(\sum_{\alpha}|\delta\bar{n}_{\alpha}|^{2})^{-2}\sum_{\alpha}|\delta\bar{n}_{\alpha}|^{4}$. For a non-interacting (interacting) particle, $P(\{\delta\bar{n}_{\alpha}\})=1$ ($P(\{\delta\bar{n}_{\alpha}\})<1$). We therefore expect that $P(\{\delta\bar{n}_{\alpha}\})$ should approach unity whenever interactions become irrelevant.
%\jp{JP: I don't think we should call this a purity, that is the square of the density matrix its a name that already has a different meaning.} \miguel{ MG: Changed name to "ocupation inverse participation ratio".}

The nature of the low-energy excitations can also be inspected by analysing the long wavelength (small $q$) behaviour of the static structure factor defined as $S(q)=N^{-1}\sum_{j,l}[\langle n_{j}n_{l}\rangle-\langle n_{j}\rangle\langle n_{l}\rangle]e^{\textrm{i}q(j-l)}$.
In the LL phase, the Luttinger liquid correlation parameter $K$ can
be computed through $K=2\pi\lim_{q\rightarrow0}S(q)/q$ \footnote{Note that the factor of 2 is needed since we are working with spinless
fermions.}\citep{Ejima2005,PhysRevB.59.4665,Ejima_2009,PhysRevLett.95.096401}. In a gapless non-interacting and translationally invariant system, it is easy to show that $K=1$. Inside an (interacting) LL phase, however, $K\neq 1$ in general.

To inspect the localization properties of the many-body wave function, we also computed inverse participation ratios (${\rm IPR}$) \citep{RevModPhys.80.1355}
 for the density fluctuations $\delta n_{i}\equiv C_{e}^{ii}$ and for the most occupied (with $\delta \bar{n}_\alpha$ closest to 1) natural
orbital that we write as $\ket{\alpha=0}=\sum_{i}\psi_{i}^{(0)}\ket i$,
where $\ket i=c_{i}^{\dagger}\ket 0$ and $\ket 0$ is the vacuum:
\begin{equation}
\begin{array}{cc}
{\rm IPR}(\{\delta n_{i}\}) & =(\sum_{i}|\delta n_{i}|)^{-2}\sum_{i}|\delta n_{i}|^{2}\\
{\rm IPR}_{{\rm NO}}(q) & =(\sum_{i}|\psi_{i}^{(0)}|^{2})^{-q}\sum_{i}|\psi_{i}^{(0)}|^{2q}.
\end{array}\label{eq:many_body_IPRs}
\end{equation}

In the non-interacting limit it is easy to show that $\delta n_{i}=|\psi_{i}^{N_{p}}|^{2}$
and $\psi_{i}^{(0)}=\psi_{i}^{N_{p}}$, where $\psi_{i}^{N_{p}}$
is the amplitude of the $N_{p}$-th single-particle wave function
(ordered by increasing eigenenergy) at site $i$. The quantities in
Eq.$\,$\ref{eq:many_body_IPRs} are therefore many-body generalizations
of the single-particle IPR \citep{Aulbach_2004} used to study the
localization properties of single-particle eigenstates. For the conventional definition of the IPR (with $q=2$ in Eq.$\,$\ref{eq:many_body_IPRs}), we have ${\rm IPR}\sim N^{-D_{2}}$, where $D_{2}$ is the fractal
dimension, with $D_{2}=1$ for delocalized states, $D_{2}=0$ for
localized states and $0<D_{2}<1$ for critical states. For the generalized version, with $q \neq 0$, we have ${\rm IPR}\sim N^{-\tau(q)}$, with $\tau(q)=D_q (q-1)$ and $D_q=d$ (with $d$ the system's dimension) for fully delocalized single-fractal states, while $D_q$ is a non-linear function of $q$ for multifractal states.

\paragraph*{Universal description around criticality.--- }

As we previously stated, Fig.$\,$\ref{fig:1} shows that the results
for the quantity $\delta E_{\kappa\varphi}$ in significantly different
models can be collapsed into a single universal scaling function.
To obtain the collapse in Fig.$\,$\ref{fig:1}(c), we first defined
the normalized distance to the critical point $v(N,\bm{\lambda})=[V-V_{c}(N,\bm{\lambda})]/V_{c}(N,\bm{\lambda})$,
where $\bm{\lambda}=(V_{2},t_{2},U,U_{2})$ contains the model parameters
(other than $V$) and $V_{c}(N,\bm{\lambda})$ is the value of $V$
at the self-dual point ($\delta E_{\kappa\varphi}(N,V,\bm{\lambda})=1$).
The reason why we use $V_{c}(N,\bm{\lambda})$ and not $V_{c}\equiv V_{c}(\infty,\bm{\lambda})$
is that for smaller systems there can be some dependence of $V_{c}(N,\bm{\lambda})$
on $N$. Such dependence can arise not only from finite-size effects,
but also because increasing $N$ also slightly modifies the filling
$\rho=N_{p}/N$ (due to $N$ being odd) and the value of $\tau_{c}$ (see table \ref{table_approxs} in SM).

Further assuming that 
\begin{equation}
    \delta E_{\kappa\varphi}\sim e^{-N/\xi}
\end{equation}
with $\xi\sim v^{-\nu}(N,\bm{\lambda})$ and extracting $\nu$ from
a fit using all the obtained data points, we get a critical exponent
$\nu=1.001\pm0.007$ (see SM). Therefore, we set $\nu=1$ and obtain
an excellent collapse shown in Fig.$\,$\ref{fig:1}(c), around the critical
point, i.e. around $v(N,\bm{\lambda})=0$. 
%We note that in principle,
%the proportionality constant that relates $\xi$ and $v^{-\nu}(N,\bm{\lambda})$
%can depend on $\bm{\lambda}$, but this dependence can be neglected
%close to criticality for the models here studied, which is confirmed
%by the obtained  collapse. 
In the SM we also show that this
collapse is not a special feature of half-filling, by also considering
the case $\rho=1/3$. We conjecture that the collapse should be observed
for any filling that is not commensurate with $\tau$ as defined in
\citep{PhysRevB.101.174203}, i.e., that does not satisfy $\rho=\mod(n\tau_{c},1)$,
with $n$ an integer that does not depend on system size. At such
commensurate fillings, single-particle gaps are opened for any strength
of the quasiperiodic potential.
%[BRUNO: are these CDW phases due to nesting?]
%[MIGUEL: Yes essentially, these are gaps that can be understood in perturbation theory by the coupling of momenta by multiples of the quasiperiodic wave vectors]

\begin{figure}[h!]
\centering{}\includegraphics[width=1\columnwidth]{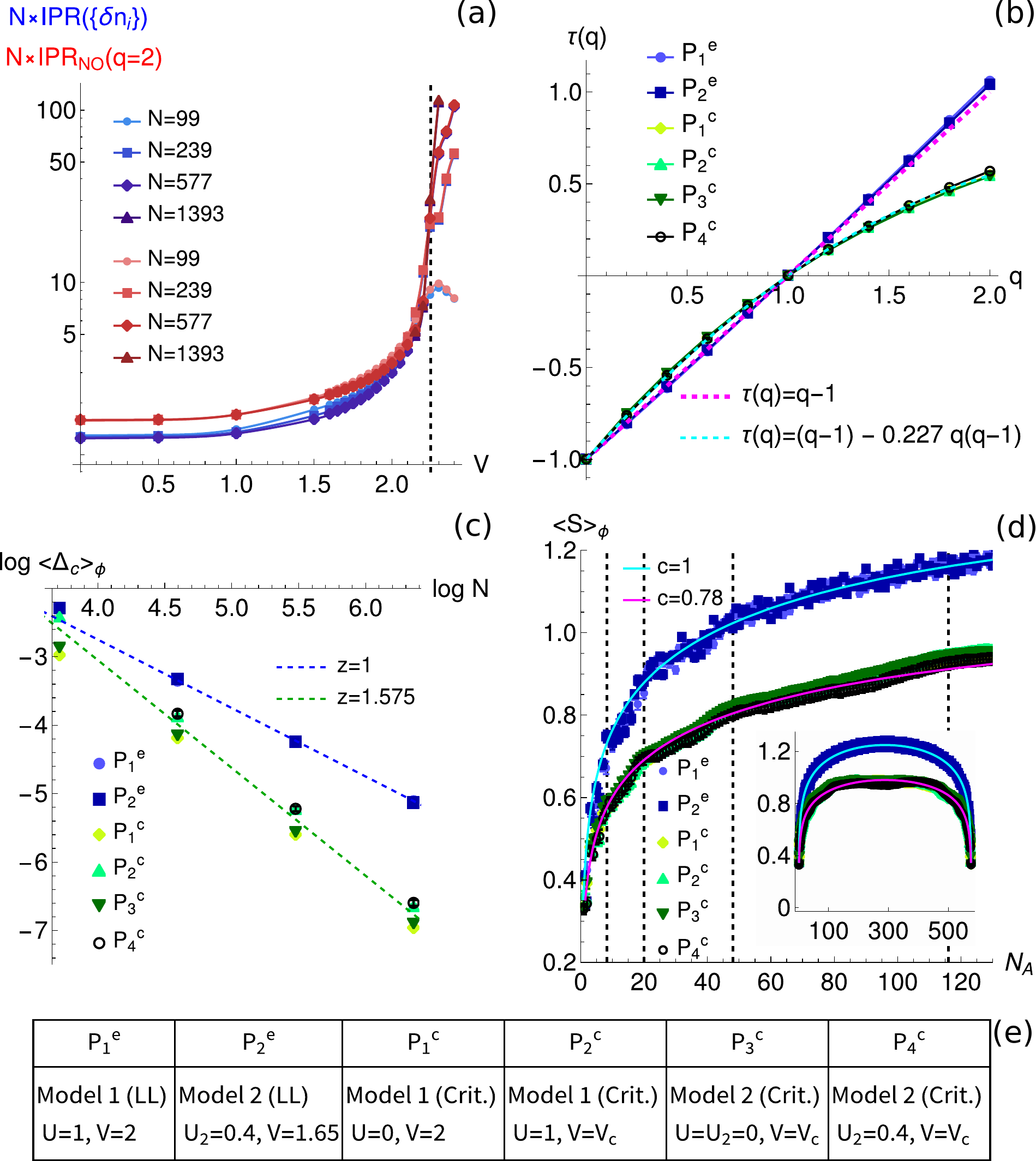}\caption{ (a) $N{\rm IPR}(\{\delta n_{i}\})$ and $N{\rm IPR}_{{\rm NO}}$ as a function of $V$ for $U=1,\phi=1.123$
and $V_{2}=t_{2}=U_{2}=0$. The vertical dashed line shows the critical
point estimated from $\delta E_{\kappa\varphi}$. (b) Exponent $\tau(q)$ defined in Eq.$\,$\ref{eq:many_body_IPRs} for different models with chosen parameters indicated in table (e), where $P_{i}^{e}$($P_{i}^{c}$) denote extended
(critical) points. In table (e), $V_{c}$ corresponds to the
estimated critical point. $\tau(q)$ was computed from linear fits to data points $(\log N, \langle {\rm IPR}_{{\rm NO}}(q) \rangle_{\phi})$, where $\langle \rangle_{\phi}$ denotes an average over different choices of $\phi$. We took $\phi_j=2\pi j/N_c, \textrm{ }j=0,\cdots,N_c-1$ and $N_c \in [100-300]$. We used system sizes $N\in \{99,239,577 \}$ for (interacting) points $P_1^e, P_2^e, P_2^c, P_4^c$ and  $N\in \{99,239,577,1393 \}$ for (non-interacting) points $P_1^c, P_3^c$.
(c,d) $\langle \Delta_{c} \rangle_{\phi}$ and $\langle  \mathcal{S}(N_{A}) \rangle_{\phi}$ computed by employing the same averaging procedure used in (b).  In (c), the blue
and green dashed lines show the scaling behaviour in the
non-interacting Aubry-André model, respectively at extended and critical points. In (d), the calculation is done for $N=577$. The main figure shows a close-up at smaller $N_A$, while the inset shows the results for all possible $N_A$. The vertical dashed lines are guides to the eye for the maxima of the log-periodic oscillations, given by $N_A^{(n)}=116/p^n$, with $n=0,1,2,3$ and $p=\lim_{m \rightarrow \infty} N_m/N_{m+1}=1+\sqrt{2}$. The magenta and cyan curves correspond to fits to Eq.$\,$\ref{eq:S_calebrese}
with $c=1$ and $c=0.78$, respectively.
%\jp{JP: Need a better label scheme then this table. Can we show the collapse of the IPRs in the inset. The log-periodic oscillations in $S$ are hard to see as plotted}
%\miguel{MG: 
%(i) Regarding the labeling scheme, what do you suggest? Putting only the parameters in the caption?
%(ii) The IPRs do not collapse perfectly in the localized phase because for that to be the case an averaging over $\phi$ would be necessary. In the LL phase however, since the $\phi$ dependence is almost neglegible, we can see a very nice collapse of $N \times IPR$. 
%(iii) Changed panel (d) to make the log-periodic oscillations clear by zooming in. Left full plot as inset, as we discussed.}
\label{fig:3}}
\end{figure}

\paragraph*{Non-interacting excitations and additional critical exponents.--- }

We have seen from the quantity $\delta E_{\kappa\varphi}$ in Eq.~\eqref{eq:DeltaEkphi}
that the effects of interactions on the scaling function and $\nu$ are irrelevant. 
We now show that particles become
effectively non-interacting in this regime and that the critical properties
obtained at different critical points are identical. For the results
that follow, we use open boundary conditions. 

In Fig.$\,$\ref{fig:2}(a), we show that the occupation inverse participation ratio approaches $1$ around the critical point. This implies that the single-particle
gapless excitations acquire a non-interacting nature. The same conclusion
can be drawn by inspecting the behaviour of the Luttinger parameter
$K$, in Fig.$\,$\ref{fig:2}(b). At small $V$, $K$ does not vary
significantly (note that when $V=0$, $K$ is known exactly for model
1 \citep{Ejima2005}). 
%\jp{JP: Give the reader some physical insight as to why} \miguel{MG: As to why $K$ doesn't vary significantly at small $V$?}
On the other hand, as $V$ gets closer to the
critical point, $K$ approaches the non-interacting value $K=1$.
Exactly at the different  critical points, the system is
no longer a LL and at small $q$ we find log-periodic corrections $S(q)\approx 0.7 q[1+a\sin(b\log q+\alpha)]$
%with $K'\approx0.7$ 
(see SM).
In the AG phase, $S(\delta q)/\delta q$, computed for the smallest non-vanishing momentum $\delta q$, decreases since $S(q)\sim q^{2}$ when $q\rightarrow0$ {[}see SM for explicit
plots of $S(q)${]}.

In Fig.$\,$\ref{fig:3}(a) we show a representative example of the
quantities $N\,{\rm IPR}(\{\delta n_{i}\})$ and $N\,{\rm IPR}_{{\rm NO}}$
across the LL-AG transition. We observe LL (AG) phase, is characterized
by ${\rm IPR}(\{\delta n_{i}\}),{\rm IPR}_{{\rm NO}}\sim N^{-1}$
($\sim{\rm const.}$), which is confirmed by the collapse (divergence)
of the curves below (above) the transition for different $N$, in
direct analogy with the results for the single-particle IPR in the
non-interacting case. Note that both quantities become almost quantitatively equal close to the critical point and at the AG phase. At the critical point we expect
multifractal scaling  with an infinite set of critical exponents (i.e. the multifractal spectrum). 
%However, in order to mitigate
%finite-size effects and obtain the exponent accurately, we average
%over different configurations of $\phi$. 
Averaging our results over $\phi$, we compute the exponent $\tau_q$ defined in Eq.$\,$\ref{eq:many_body_IPRs}, that we show in Fig.$\,$\ref{fig:3}(b). In this figure, we can see that in the LL phase ($P_1^e$ and $P_2^e$), $D_q \approx 1$, while at interacting critical points ($P_2^c$ and $P_4^c$) we observe a multifractal behaviour quantitatively compatible with the one obtained at the non-interacting critical points ($P_1^c$ and $P_3^c$), that is, $D_q \approx 1 - 0.227 q$ \footnote{By fitting to the behaviour $D_q = c_1 + c_2 q $, we obtained $c_1 = 0.992 \pm 0.005$ and $c_2 = -0.227  \pm 0.004$ at the critical point of the non-interacting Aubry-André model.}. 
In the SM, we show explicit data for 
$ \langle {\rm IPR}_{{\rm NO}}(q) \rangle_{\phi}$ as a function of $N$, from which the exponents $\tau(q)$ were extracted.

We also computed the $\phi$-averaged scaling of the charge gap $\Delta_{c}$ in Fig.$\,$\ref{fig:3}(c), that allowed us to extract dynamical critical exponents $z$. Remarkably,
the scaling exponents are compatible with the exponents obtained for
the non-interacting Aubry-André model. This, together with the multifractal analysis, is a strong indication that the universality
class of the delocalization-localization transition is unchanged upon
the addition of interactions. An important remark is that, as seen
in Ref.$\,$\citep{PhysRevB.101.174203}, the dynamical exponent for
the non-interacting Aubry-André model can depend on $\rho$ and $\tau$. Since here
we are fixing the latter, a natural question is whether the independence
of the critical exponents on the model is a special feature of our
choice. In the SM, we argue that this is not the case by obtaining
compatible finite-size scalings of the charge gap at critical points
of different non-interacting models for other choices of $\rho$ and
$\tau$.

Finally, we also plot the $\phi$-averaged entanglement entropy as
a function of the size of bipartion $A$, $N_{A}$, in Fig.$\,$\ref{fig:3}(d).
In the LL phase, $S$ follows the behaviour of Eq.$\,$\ref{eq:S_calebrese}
with $c=1$, as in the non-interacting delocalized case \citep{PhysRevA.87.043635}.
At the critical point, the results are compatible with the non-interacting
Aubry-André model result, showing corrections to Eq.$\,$\ref{eq:S_calebrese} in the form of log-periodic oscillations,
similarly with what was observed for critical aperiodic spin chains
in Ref.$\,$\citep{Igloi_2007} (see SM for more detailed analysis of the log-periodic oscillations). A fit to Eq.$\,$\ref{eq:S_calebrese}
neglecting these corrections yields $c\approx0.78$, in agreement
with \citep{PhysRevB.102.064204} (note, however, that in this case
$c$ cannot be interpreted as a central charge).

\begin{figure}[h]
\centering{}\includegraphics[width=1\columnwidth]{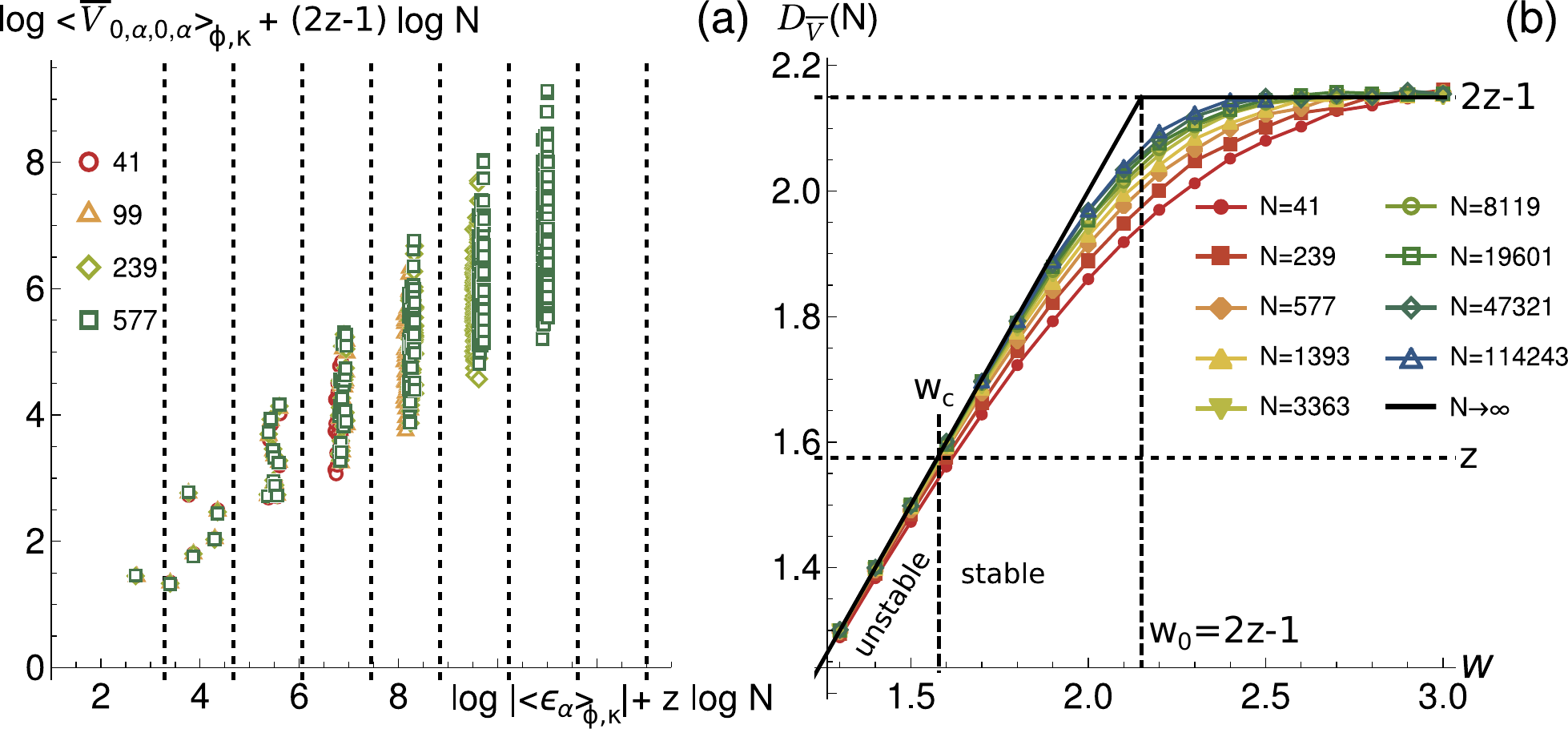}\caption{
{\bf Scaling dimension of long range interactions}.
(a) Collapses of $\langle \bar{V}_{0\alpha0\alpha} \rangle_{\phi,\kappa}$ calculated at the critical point of the half-filled Aubry-André model ($V=2$) for $\tau=1/\sqrt{2}$ (that has $z=1.575$), for different system sizes and different energies, and for power-law interactions with decay exponent $w=3$. Twisted boundary conditions are used and the results are averaged over $200$ random configurations of $\phi$ and $\kappa$, denoted by $\langle \rangle_{\phi,\kappa}$. (b) Finite-size results for $D_{ \bar{V} }(N,w)$. The dashed lines indicate $D_{ \bar{V} }=z$ and $D_{ \bar{V} }=2z-1$. The black lines indicate the expected thermodynamic-limit behaviour. $w_c$ corresponds to the critical value of the exponent $w$ below which interactions become relevant. 
\label{fig:4}
}
\end{figure}

\paragraph{Generalized Chalker scaling and irrelevance of generic short-range interactions.--- } 
We now provide a framework to understand why the short range interactions we have studied so far are irrelevant. Our argument relies on a tree-level scaling analysis of the interaction at the critical point of the Aubry-Andr\'e model. For completeness, we extend our discussion to long-range interactions, of the form $r^{-w}$ with $w>1$, and show there is a critical power law, $w_c$, where they eventually become relevant.
%In order to increase the generality of our results to more generic interactions than those of model in Eq$\,$\ref{eq:H}, here we show that generic short-range interactions are irrelevant at the critical point of the Aubry-André model in the $U\rightarrow 0$ limit. Interestingly, we also show that long-range interactions can become relevant if they decay slowly enough. 
We employ twisted boundary conditions and choose the long-range interaction to be a periodized form of the power-law potential $U \sum_{r,j} j^{-w} n_{r}n_{r+j}$, given by   $U\sum_{j=1}^{N-1}N^{-w}\zeta(w,j/N)\sum_{r=1}^N n_{r}n_{r+j} $, where $\zeta(w,y)= \sum_{k=0}^{\infty} (k+y)^{-w}$ is the Hurwitz zeta function and $n_{r}=n_{r+N}^{\dagger}$ due to twisted boundary conditions. To compute the scaling dimension of the interaction term, denoted as $D_U(w)$,
we write the interacting Hamiltonian on the single-particle eigenbasis of the non-interacting Aubry-André model Hamiltonian, $H_0$. 
We label single-particle states with Greek indices, ${\ket{\alpha}}$, and single-particle energies by $\epsilon_{\alpha}$. In this base 

\begin{equation}
H = \sum_{\alpha} \epsilon_{\alpha} c^{\dagger}_{\alpha} c_{\alpha} -  \sum_{\alpha,\beta,\gamma,\delta}\bar{V}_{\alpha\beta\gamma\delta}c^{\dagger}_{\alpha}c^{\dagger}_{\beta} c_{\gamma}c_{\delta}
\end{equation}

\noindent where $\bar{V}_{\alpha\beta\gamma\delta}$ is the antisymmetrized version of the interaction tensor in the eigenbasis of the Aubry-Andr\'e
model
\begin{equation}
    V_{\alpha\beta\gamma\delta}=\frac{U}{N^w}\sum_{r=1}^N \sum_{j=1}^{N-1}\zeta\Big(w,\frac{j}{N}\Big) \braket{\alpha|r}\braket{\beta|r+j}\braket {r|\gamma}\braket{r+j|\delta},
    \label{Vabgd}
\end{equation}
 with $\braket{\alpha|r} = \bra{0} c_{\alpha} c^\dagger_r \ket{0}$.
%As a result, the scaling dimension of the interaction tensor denoted as $D_V(w)$, and defined through $ \bar{V}_{0\alpha0\alpha} \rightarrow N^{D_{ \bar{V}}(w)}\bar{V}_{0\alpha0\alpha}$ determines the scaling dimension of $U$.

The leading contributions to the interacting term come from states with energies around the Fermi level, $E_F$. In the following, we denote by $\epsilon_0$ the energy closest to $E_F$ and we set $\epsilon_0=0$ for convenience. 
%If all the indices are fixed at the Fermi level, we have by antisymmetry  that $\bar{V}_{0 0 0 0}=0$. By the same argument, a vanishing contribution is also obtained if only one of the indices is allowed to vary, e.g. $\bar{V}_{\alpha 0 0 0}$. 
By the antisymmetry of $V$, the lowest-order non-vanishing contributions involve setting two indices to the Fermi level and varying the remaining, i.e. $\bar{V}_{\alpha 0 \beta 0}$. Among those, we find that the dominant contribution arises for $\alpha=\beta$ (see SM) and thus we may restrict our analysis to an interaction tensor of the form   $\bar{V}_{0\alpha0\alpha}$ for small $|\epsilon_{\alpha}|$.
%$|\epsilon_{\alpha}-\epsilon_0|$. 
%We can check numerically that this term is self-similar 

For the chosen model (half-filling, with $\tau=1/\sqrt{2}$), the critical point of the (non-interacting) Aubry-Andr\'e model under a \emph{discrete} scale transformation $r \to N r$ is invariant under the rescaling $\epsilon_{\alpha} \rightarrow N^z \epsilon_{\alpha}$. 
The interacting term transforms as $\bar{V}_{0\alpha0\alpha} \rightarrow N^{D_{ \bar{V}}(w)}\bar{V}_{0\alpha0\alpha}$, where $D_{ \bar{V}}$ is the scaling dimension of the interaction tensor that can be obtained by the data collapse illustrated in Fig.~\ref{fig:4}(a). In this example, we take the  half-filled Aubry-Andr\'e model with $\tau=1/\sqrt{2}$, that has $z=1.575$, and set $w=3$, finding that $D_{ \bar{V}}(w=3)=2z-1$. 
%At the critical point of the (non-interacting) Aubry-Andr\'e model the interaction tensor has \emph{discrete} scale invariance $r \to N r$, under the rescaling $\epsilon_{\alpha} \rightarrow N^z \epsilon_{\alpha}$ and $ \bar{V}_{0\alpha0\alpha} \rightarrow N^{D_{ \bar{V}}(w)}\bar{V}_{0\alpha0\alpha}$, where $D_{ \bar{V}}$ is the scaling dimension of the interaction tensor that can be obtained by data collapse as shown in Fig.~\ref{fig:4}.
The relation between the energy, $\epsilon_{\alpha}$, and the interaction strength, $\bar{V}_{0\alpha0\alpha}$, follows a generalized Chalker scaling \citep{PhysRevLett.61.593,CHALKER1990253,PhysRevB.76.235119,PhysRevB.89.155140,PhysRevB.101.235121}.
However, a significant difference to previous Chalker scaling analyses is the full antisymmetrization of the interaction term that follows from fermionic statistics.
%We point out, however, the importance of antisymmetrizing the interaction term in order to correctly obtain its scaling dimension. 
%Studying the non-antisymmetrized tensor in Eq.$\,$\ref{Vabgd} yields quantitatively and qualitatively different results, therefore invalidating that approach.  
%
%Since the interaction tensor measures correlations between two single-particle eigenstates with energies $\epsilon_0$ (Fermi level) and $\epsilon_{\alpha}$, the scaling collapse implies that it follows a generalized Chalker scaling \citep{PhysRevLett.61.593,CHALKER1990253,PhysRevB.76.235119,PhysRevB.89.155140,PhysRevB.101.235121}. We point out, however, the importance of antisymmetrizing the interaction term in order to correctly obtain its scaling dimension. Studying the non-antisymmetrized tensor in Eq.$\,$\ref{Vabgd} yields quantitatively and qualitatively different results, therefore invalidating that approach.  
%

By power-counting, we find the scaling dimension of the interaction to be $D_U = z-D_{ \bar{V}}$, implying that interactions are irrelevant if $D_{ \bar{V}}>z$ (see SM for details). To infere $D_{ \bar{V}}(w)$ 
in the thermodynamic limit, we studied the finite-size dimension $D_{\bar{V}}(w,N_{m})$ (where $m$ labels the order of the approximant size $N_m$), which satisfies $D_{\bar{V}}(w,\infty) \equiv D_{\bar{V}}(w)$, and can be computed through  
$D_{\bar{V}}(w,N_{m})=-(\log N_{m+1}-\log N_{m})^{-1} 
[ \log\bar{V}^{m+1}_{0 1 0 1}(w)-\log \bar{V}^m_{0 1 0 1}(w) ] $, as depicted in Fig$\,$\ref{fig:4}(b). As for the case $w=3$ shown in  Fig$\,$\ref{fig:4}(a), for sufficiently large $w>w_0=2z-1$, $D_{ \bar{V}}=2z-1$. This scaling is also retrieved for other types of short-range interactions (e.g. finite range or exponentially suppressed), as we show in detail in the SM. Since $z>1$ at the critical point, interactions are always irrelevant in this case. This justifies the findings of previous sections near $U=U_2=0$. For $w<w_0$, the finite-size results shown in Fig\,\ref{fig:4}(b) are compatible with $D_{ \bar{V}}=w$. In this case, interactions become relevant for $w<z$ since at that point we start having $D_{ \bar{V}}<z$ and thus $D_U>0$. The nature of this interesting fixed-point is left for future exploration.

\paragraph{Discussion.--- }

For a broad class of quasiperiodic models, we provided solid evidence
that (i) short-range interactions are irrelevant at the LL-AG transition, not affecting the
non-interacting critical exponents; (ii) a many-body generalization
of the theory proposed in \citep{GoncalvesRG2022} can be formulated;
 and (iii) in the limit of vanishing interactions, the non-interacting critical point is robust to any short-range (and even some long-range) interactions. Our work not
only provides a unified understanding of LL-AG transitions around criticality
in terms of flows to non-interacting fixed-points accompanied by the
emergence of many-body dualities in widely different models, but it
also offers a very precise way to estimate the critical points. Future interesting questions to address include the effect of interactions on critical phases of the non-interacting quasiperiodic models, see e.g. the models in  \citep{Liu2015,anomScipost,critical_phase_goncalves}, and the nature of the fixed-point at which long-range interactions become relevant at the non-interacting Aubry-André critical point. 

\label{sec:Discussion}
\begin{acknowledgments}
The authors MG and PR acknowledge partial support from Fundação para
a Ciência e Tecnologia (FCT-Portugal) through Grant No. UID/CTM/04540/2019.
BA and EVC acknowledge partial support from FCT-Portugal through Grant
No. UIDB/04650/2020. MG acknowledges further support from FCT-Portugal
through the Grant SFRH/BD/145152/2019. BA acknowledges further support
from FCT-Portugal through Grant No. CEECIND/02936/2017. 
JHP is patially supported by the Air Force Office of Scientific Research under Grant No.~FA9550-20-1-0136, and the Alfred P. Sloan Foundation through a Sloan Research Fellowship.
 We finally acknowledge the Tianhe-2JK cluster at the Beijing Computational Science
Research Center (CSRC), the Bob\textbar Macc supercomputer through
computational project project CPCA/A1/470243/2021 and the OBLIVION
supercomputer, through projects HPCUE/A1/468700/2021, 2022.15834.CPCA.A1 and 2022.15910.CPCA.A1
(based at the High Performance Computing Center - University of Évora)
funded by the ENGAGE SKA Research Infrastructure (reference POCI-01-0145-FEDER-022217
- COMPETE 2020 and the Foundation for Science and Technology, Portugal)
and by the BigData@UE project (reference ALT20-03-0246-FEDER-000033
- FEDER and the Alentejo 2020 Regional Operational Program. Computer
assistance was provided by CSRC's, Bob|Macc's and OBLIVION's support
teams. 
% See https://rnca.fccn.pt/projetos/ for project names

\end{acknowledgments}

\bibliographystyle{apsrev4-1}
\bibliography{Quasiperiodic_Interactions,Entanglement_entropy_Conformal_field_theory,Non_interacting_refs}

%%%%%%%%%%%%%%%%%%%%%%%%%%%%%%%%%%%%%%%%%%%%%%%%%%%%%%%%%

%%%%%%%%%%%%%%%%%%%%%%%%%%%%%%%%%%%%%%%%%%%%%%%%%%%%%%%%%

%%%%%%%%%%%%%%%%%%%%%%%%%%%%%%%%%%%%%%%%%%%%%%%%%%%%%%%%%

\clearpage\onecolumngrid

\beginsupplement
\begin{center}
\textbf{\large{}Supplemental Material for: \vspace{0.1cm}
}{\large\par}
\par\end{center}

\begin{center}
{\large{}Short-range interactions are irrelevant at
the quasiperiodic-driven Luttinger Liquid to Anderson Glass transition}{\large\par}
\par\end{center}

\vspace{0.3cm}

\tableofcontents{}

\newpage

\section{System size approximants used in finite-size simulations}

In our finite-size simulations, we use rational approximants of $\tau$,
$\tau_{c}=p/N$, with $p$ and $N$ co-prime numbers. These approximants
were chosen to be exact convergents of the continued fraction expansion
of $\tau$. This can be done as long as the unit cell defined by $\tau_{c}$
is equal to or larger than the system size, which guarantees that
the system remains incommensurate. For our choice, the size of the unit cell is exactly the system size $N$. 
%This choice also enables the application of
%periodic boundary conditions without defects. 
We chose the series of approximants given in table \ref{table_approxs}.

\begin{center}
\begin{table}[h!]
\begin{tabular}{|c|c|c|c|c|c|c|c|}
\hline 
$N$ & 17 & 41 & 99 & 239 & 577 & 1393 & $\infty$\tabularnewline
\hline 
\hline 
$\tau_{c}$ & $\frac{12}{17}$ & $\frac{29}{41}$ & $\frac{70}{99}$ & $\frac{169}{239}$ & $\frac{408}{577}$ & $\frac{985}{1393}$ & $\frac{1}{\sqrt{2}}$\tabularnewline
\hline 
\end{tabular}.
\caption{System size approximants of $\tau=1/\sqrt{2}$ used for the finite-size calculations.}
\label{table_approxs}
\end{table}
\par\end{center}

\section{Additional scaling collapses: extracting $\nu$ and going away from
half-filling}

\begin{figure}[h!]
\begin{centering}
\includegraphics[width=0.8\columnwidth]{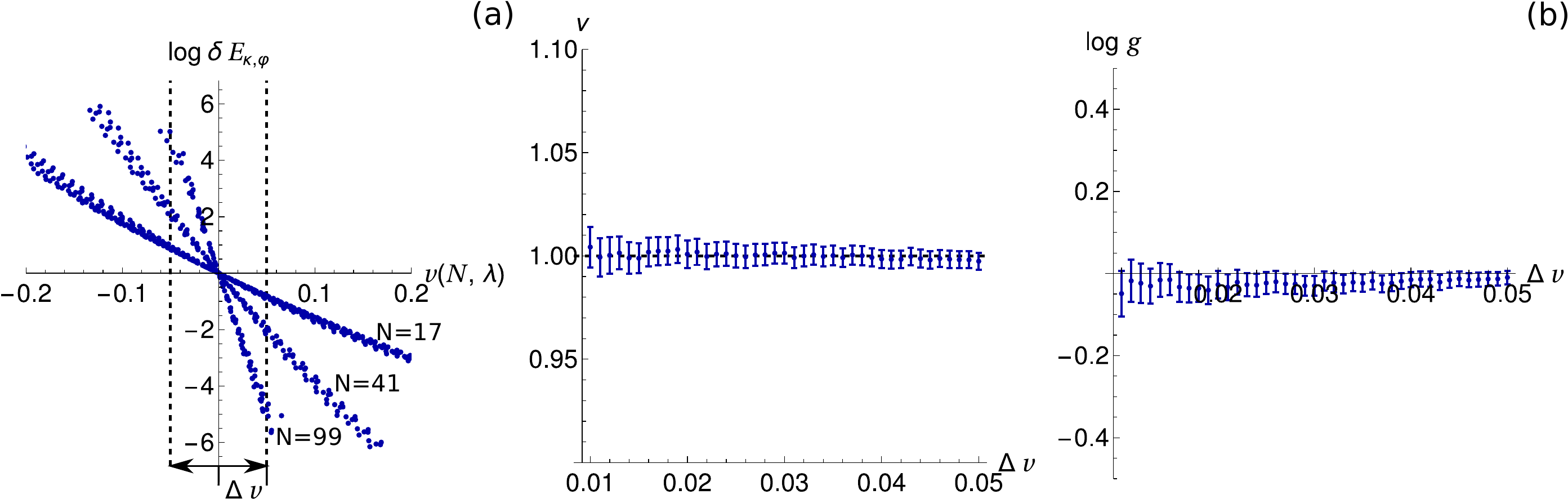}
\par\end{centering}
\caption{(a) Raw data used for fits to extract $\nu$. Each cluster of points
corresponds to a different system size, indicated close to it. (b)
Parameters extracted from a linear multivariate fit to the model $\log|\Lambda_{\kappa\varphi}|=\log N+\nu\log|v|-\log g$,
by using data points $(\log N,\log v,\log|\Lambda_{\kappa\varphi}|)$
selected for different windows $\Delta v$ {[}represented in (a){]}.
\label{fig:extracting_nu}}
\end{figure}

We start by extracting the critical exponent $\nu$ from the raw data
on $\delta E_{\kappa\varphi}$, to validate our choice of $\nu=1$
in the main text. Assuming the ansatz $\delta E_{\kappa\varphi}\propto e^{-N/\xi}$
and that $\xi=gv^{-\nu}(N,\bm{\lambda})$, we have $\Lambda_{\kappa\varphi}=\log\delta E_{\kappa\varphi}=-Nv^{\nu}(N,\bm{\lambda})/g$
(note that $\delta E_{\kappa\varphi}=1$ for $v=0$) and therefore,
we have $\log|\Lambda_{\kappa\varphi}|=\log N+\nu\log v-\log g$.
We therefore carry out a linear multivariate fit using the data points
$(\log N,\log|v|,\log|\Lambda_{\kappa\varphi}|)$ to extract $\nu$
and $\log g$. The results are in Fig.$\,$\ref{fig:extracting_nu},
where we show the fitting results as a function of the range $|\Delta v|$
below which data points were selected. The final results $\nu=1.001\pm0.007$
and $g=0.972\pm0.033$ were obtained by averaging the results (and
fitting errors) for $\nu$ and $g$, for all the considered windows
$\Delta v$. 

For the non-interacting Aubry-André model, we have that $|\xi|=|1/\log(2t/V)|$
and therefore for $v\rightarrow0$ we have $|\xi|=|v|^{-1}$ and
$|\log\delta E_{\kappa\varphi}|=-N|v|$. This is consistent with the
fitting results obtained for $\nu$ and $g$, which implies that close
enough to the critical point, the correlation length behaves in the
same way, irrespective of the considered model. Note that in principle,
$g$ could depend on $\bm{\lambda}$ (the remaining parameters of
the model), but we observed here for the studied models that close
enough to criticality, $g\approx1$.

\begin{figure}[h!]
\begin{centering}
\includegraphics[width=0.5\columnwidth]{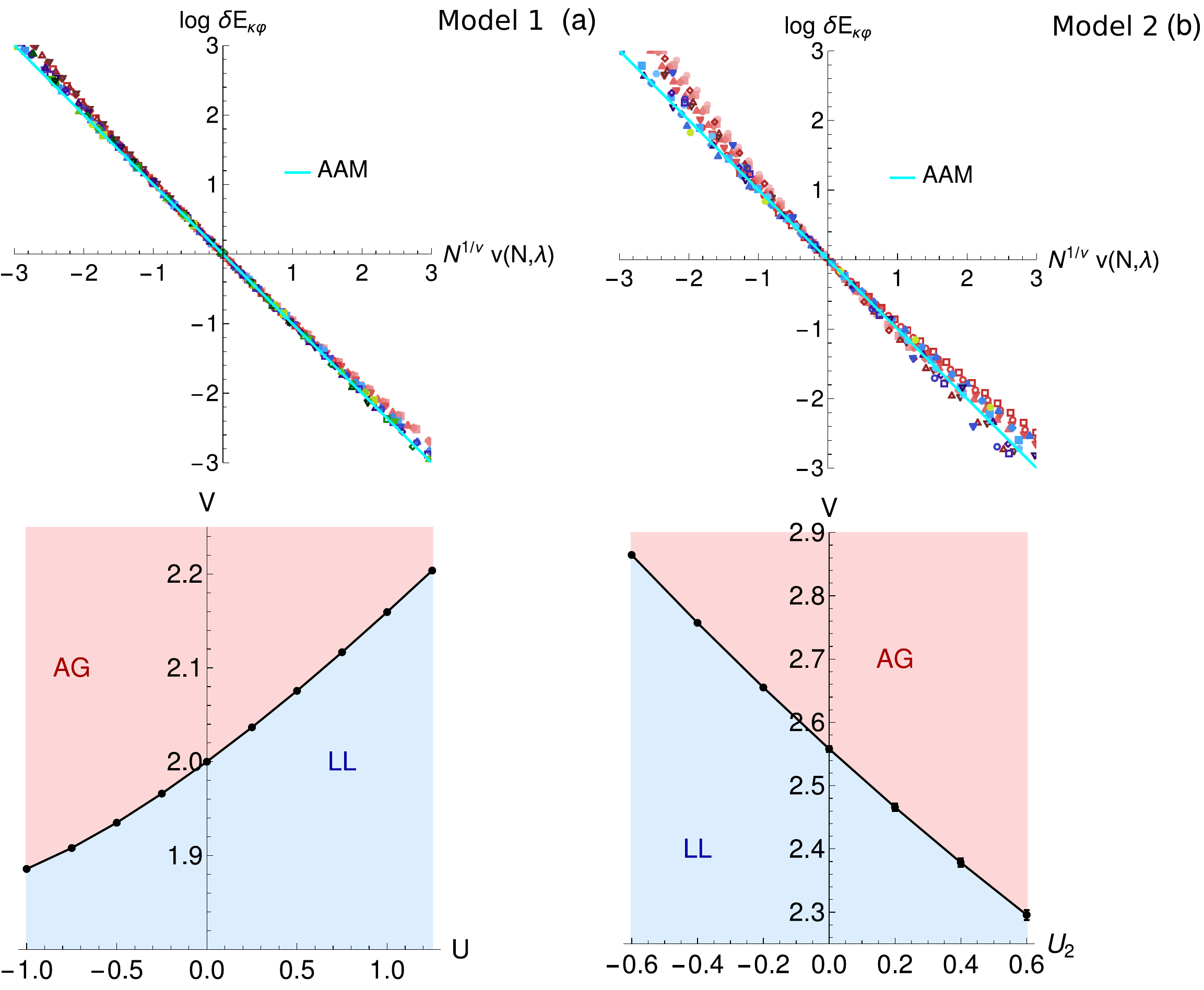}\caption{Results at filling $\rho=1/3$ for model 1 (a) and model 2 (b) defined
in the main text. The bottom panels contain the phase transition points
that we take to be $V_{c}(N=99,\bm{\lambda})$ (where $\bm{\lambda}$
contains the model parameters) with an error computed through $|V_{c}(99,\bm{\lambda})-V_{c}(41,\bm{\lambda})|$
(the difference in estimates for the largest used system sizes). Since
the error is very small, $V_{c}(N=99,\bm{\lambda})$ already provides
a very accurate estimation of $V_{c}(N=\infty,\bm{\lambda})$.\label{fig:rho1o3}}
\par\end{centering}
\end{figure}

We finally show that the 
data collapse here observed is not
a special feature of half-filling. For that purpose, we also obtain
results for a filling $\rho=1/3$, again using models 1 and 2 defined
in the main text. The results are in Fig.$\,$\ref{fig:rho1o3}, showing
nice collapses around criticality.

\section{Additional results for open boundary conditions}

\subsection{Structure factor}

We have seen in the main text that the Luttinger parameter $K$ approaches
$1$ in the Luttinger liquid phase close to criticality, which implies
that the small-$q$ behaviour of the static structure factor $S(q)$
is that of a non-interacting system. Here we explore in more detail
the $S(q)$ behaviour at the critical point and in the localized phase.
We will do so in the non-interacting (using the single-particle Hamiltonian)
and interacting (using DMRG) cases. Let us derive an expression for
$S(q)$ in the former case, using the single-particle eigenstates.
In the non-interacting case, one can easily show that 

\begin{equation}
S(q)=\frac{1}{N}\sum_{i,j=1}^{N}[(\Phi\Phi^{\dagger})_{ii}\delta_{ij}-(|\Phi\Phi^{\dagger}|^{2})_{ij}]e^{\textrm{i}q(i-j)}
\end{equation}
where $\Phi$ is a matrix containing the occupied single-particle
eigenstates in its columns and $|.|^{2}$ squares all entries of matrix
$\Phi\Phi^{\dagger}$. 

\begin{figure}[h!]
\centering{}\includegraphics[width=1\columnwidth]{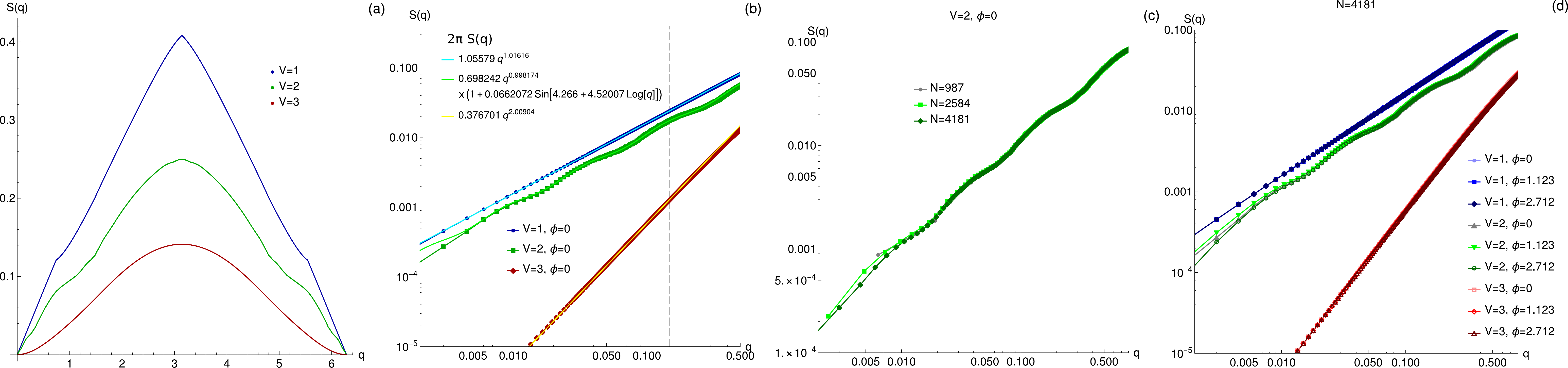}\caption{Results for non-interacting Aubry-André model, at half-filling {[} $N_{p}=\lfloor\rho N\rfloor$,
where $\lfloor x\rfloor$ takes the integer part of $x$ and $\rho=1/2$
{]}, using open boundary conditions. (a) Results for $V=1$ (extended
phase), $V=2$ (critical point) and $V=3$ (localized phase). (b)
Low $q$ behaviour of $S(q)$. The vertical dashed line denotes the
largest $q$ considered for the fits to the expressions: (i) $2\pi S(q)=Kq^{\eta}$
in the extended phase, where we obtained $K,\eta\approx1$ ; (ii)
$2\pi S(q)=Kq^{\eta}(1+a\sin[b+c\log(q)])$ in the critical phase,
where we extracted $K\approx0.7$ and $\eta\approx1$; (iii) $S(q)\sim q^{\eta}$
in the localized phase, where we extracted $\eta\approx2$. Note that
at the critical point, there are clear log-periodic oscillations.
These are not a finite-size effect, as can be seen in (c), where different
system sizes were considered and the oscillations are robust. (d)
Dependence of $S(q)$ on $\phi$, for fixed $N=4181$. The results
for different choices or $\phi$ are essentially the same, except
at the critical point for small $q$, where there is a slight $\phi$-dependence,
for fixed $N$.\label{fig:Sq_non:interacting}}
\end{figure}

In Fig.$\,$\ref{fig:Sq_non:interacting} we present results for the
non-interacting Aubry-André model. We see that at small $q$, (i) $2\pi S(q)=Kq$
and $K\approx1$ in the extended phase; (ii) $2\pi S(q)=Kq(1+a\sin[b+c\log(q)])$
and $K\approx0.7$ at the critical point; (iii) $S(q)\sim q^{2}$
in the localized phase. Interestingly, at the critical point, there
are clear $\log(q)$-periodic oscillations. 

\begin{figure}[h!]
\centering{}\includegraphics[width=0.9\columnwidth]{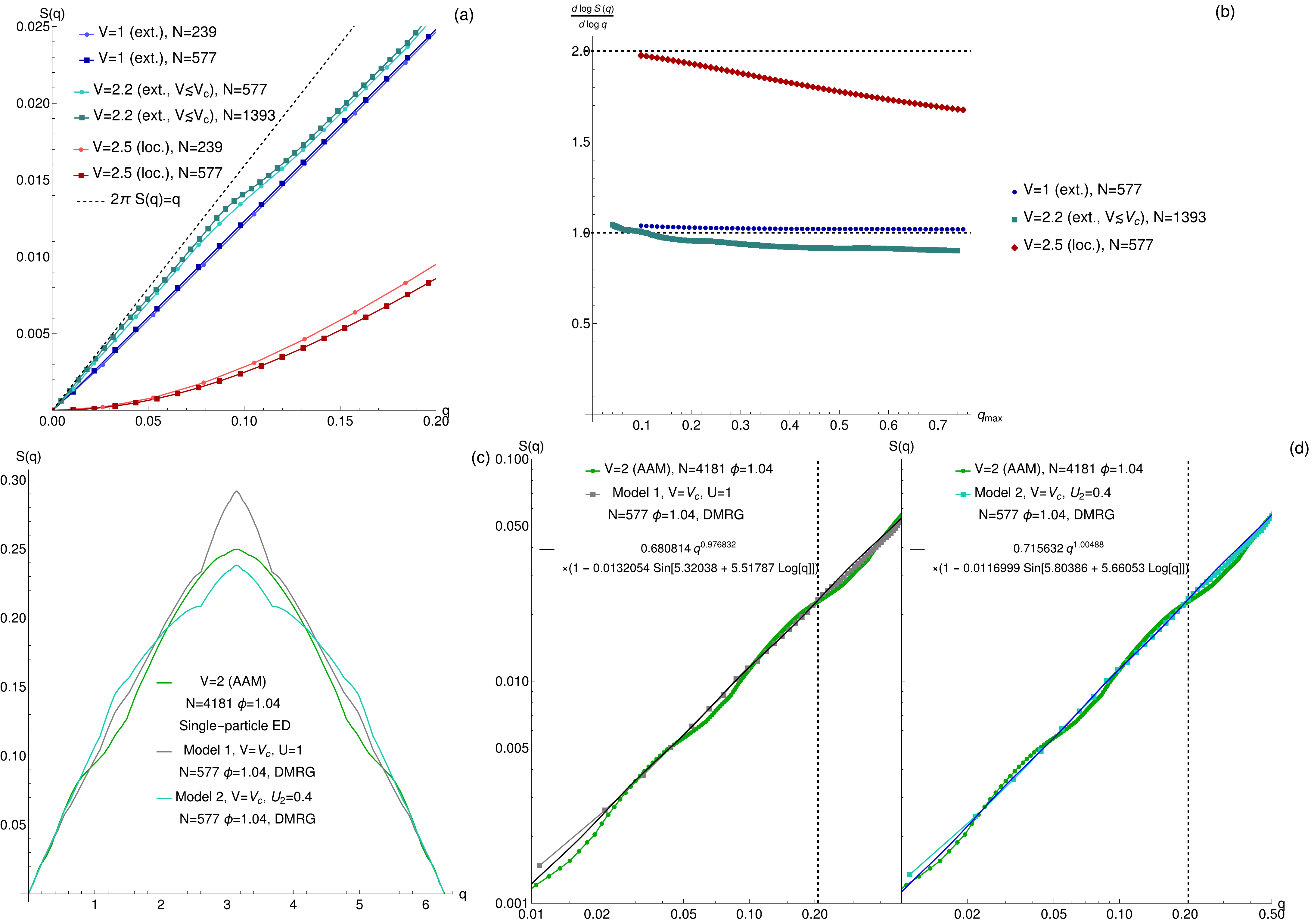}\caption{Results for interacting models, at half-filling, using open boundary
conditions. (a) $S(q)$ for $U=1$ and $U_{2}=t_{2}=V_{2}=0,\phi=1.123$,
for $V$ in the LL/extended phase ($V=1$), in the LL phase but close
to the critical point ($V=2.2$) and in the AG phase ($V=2.5$). (b)
Assuming that $S(q)\sim q^{\eta}$, we extract $\eta\equiv\protect\partial\log S(q)/\protect\partial\log q$
by making a linear fit to the $\log S(q)$ vs. $\log q$ data, from
$q=0$ up to $q=q_{{\rm max}}$ (given in the x-axis of the figure).
We see that in the LL phase, $\eta\approx1$, while in the AG, $\eta\rightarrow2$,
as in the non-interacting case. (c) $S(q)$ at critical points obtained
with significantly different parameters indicated in the figure, including
the non-interacting case. We can
see that in all cases, the small $q$ behaviour is very similar. (d) Log-log plot for the data in (c), along with fits to the expression $2\pi S(q)=Kq^{\eta}(1+a\sin[b+c\log(q)])$ for the interacting critical points, with the fit parameters given in the figure. Note that the vertical dashed line in the
right panel denotes the largest $q$ considered for the fits.  Neglecting the log-periodic oscillations, we get $2\pi S(q)=Kq$, with $K \approx 0.7$ in all cases. \label{fig:Sq_interacting}}
\end{figure}

We now consider the family of interacting models given by Eq.$\,$\ref{eq:H}
in the main text. The results for different choices of these interacting
models are given in Fig.$\,$\ref{fig:Sq_interacting}. There we see
that in the LL phase we still have $S(q)\sim q$ {[}Figs.$\,$\ref{fig:Sq_interacting}(a,b){]}.
However, we have that $2\pi S(q)=Kq$, with $K\neq1$ sufficiently
away from the critical point since the system becomes a truly interacting
LL, as in the $V=0$ limit. As the critical point is approached, we
have $K\rightarrow1$. Exactly at the critical point, on the other
hand, $S(q)$ shows an identical behaviour as in the non-interacting
Aubry-André model's critical point, see Fig.$\,$\ref{fig:Sq_interacting}(c).
It is remarkable to see that even though there are significant differences
for larger $q$ for the different considered (interacting and non-interacting)
critical points, the small-$q$ behaviour is the same. Interestingly, the amplitude of the log-periodic oscillations decreases in the interacting critical points, as can be seen in Fig.$\,$\ref{fig:Sq_interacting}(d). Finally, in the AG phase
we have $S(q)\sim q^{\eta}$ with $\eta\rightarrow2$, compatible
with the behaviour in the non-interacting localized phase, as shown
in Figs.$\,$\ref{fig:Sq_interacting}(b).

\subsection{Natural orbitals}

\begin{figure}[h!]
\centering{}\includegraphics[width=1\columnwidth]{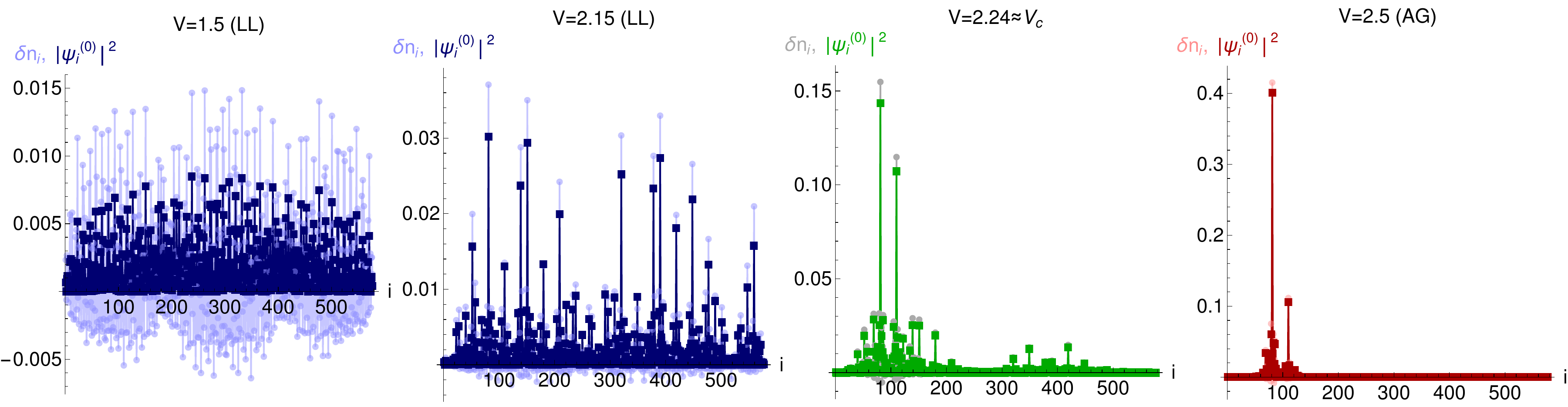}\caption{ Plots of the squared amplitudes of natural orbitals, $|\psi_{i}^{(0)}|^2$, and of the density fluctuations $\delta n_i$ for $N=577$, with $U=1$ and $U_{2}=t_{2}=V_{2}=0$, and $\phi=1.123$, for different $V$. \label{fig:No_vs_dni}}
\end{figure}

In the main text we have shown that the highest occupied natural orbitals are extended and localized, respectively at the LL and AG phases, and critical at the critical point. Here we show explicit plots, comparing the results with the density fluctuations $\delta n_i \equiv C^{ii}_e$. 
The results are in Fig.$\,$\ref{fig:No_vs_dni}. We can see that when the critical point is approached from the LL phase, $\delta n_i$ becomes very close to $|\psi_{i}^{(0)}|^2$, signaling the irrelevance of interactions (in the non-interacting case, these quantities are equal).

\begin{figure}[h!]
\centering{}\includegraphics[width=0.55\columnwidth]{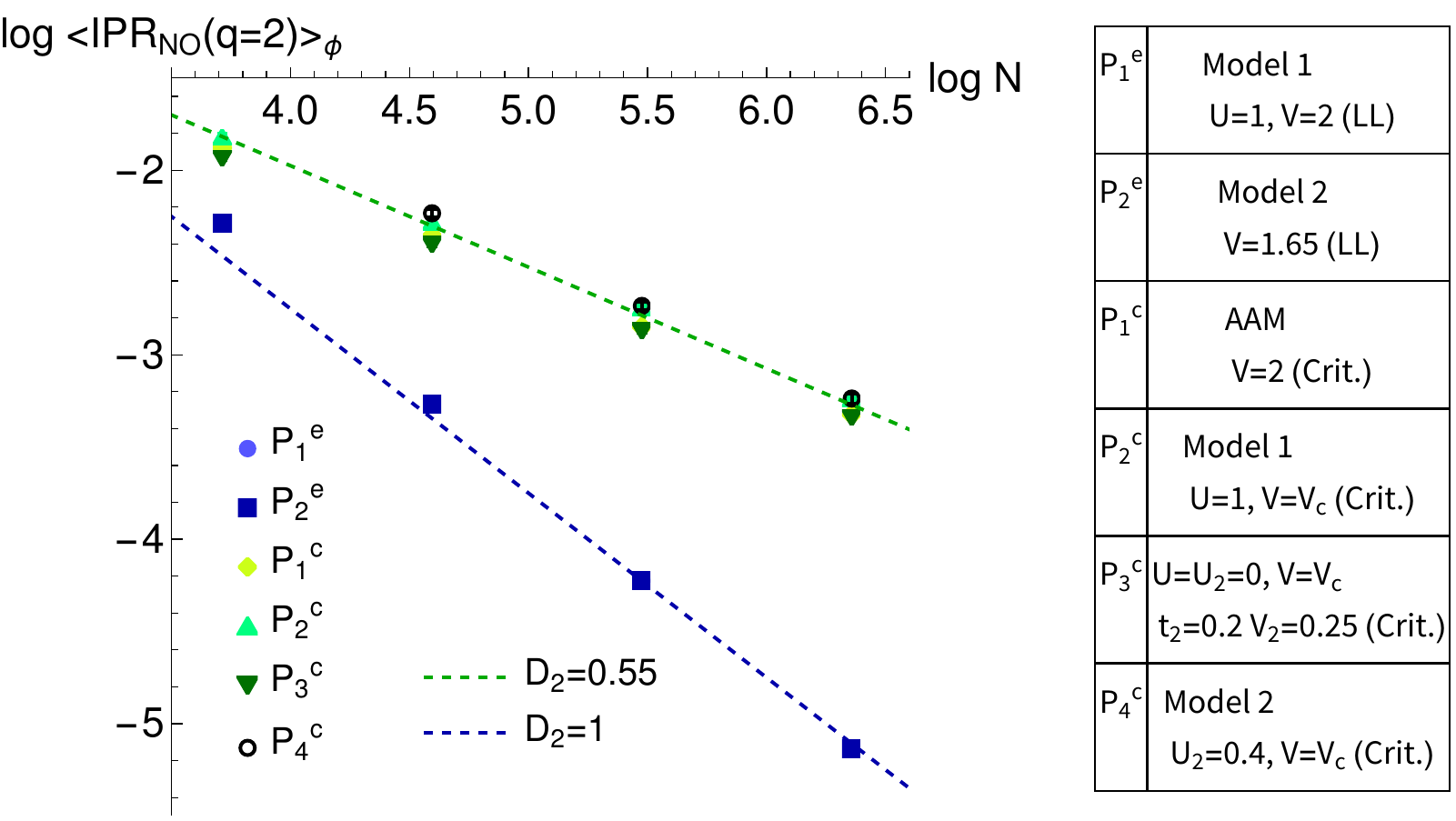}\caption{ 
 $\langle \textrm{IPR}_{N0}(q=2) \rangle_{\phi}$ for different models with chosen parameters indicated in the table at right, where $P_{i}^{e}$($P_{i}^{c}$) denote extended
(critical) points. In the table, $V_{c}$ corresponds to the
estimated critical point. $\langle \rangle_{\phi}$ denotes an average over different choices of $\phi$. We took $\phi_j=2\pi j/N_c, \textrm{ }j=0,\cdots,N_c-1$ and $N_c \in [100-300]$. The blue
and green dashed lines in (c) shows the scaling behaviour in the
non-interacting Aubry-André model, respectively at extended and critical points.
\label{fig:frac_dim}}
\end{figure}

To finish this section and complement multifractal analysis carried out in Fig.$\,$\ref{fig:3}(b) of the main text, we show explicit data for $\langle \textrm{IPR}_{N0}(q=2) \rangle_{\phi}$ as a function of system size $N$, from which the exponent $\tau(q=2)$ was extracted. The results are shown in Fig.$\,$\ref{fig:frac_dim}.

\subsection{Entanglement entropy}

\begin{figure}[h!]
\centering{}\includegraphics[width=0.85\columnwidth]{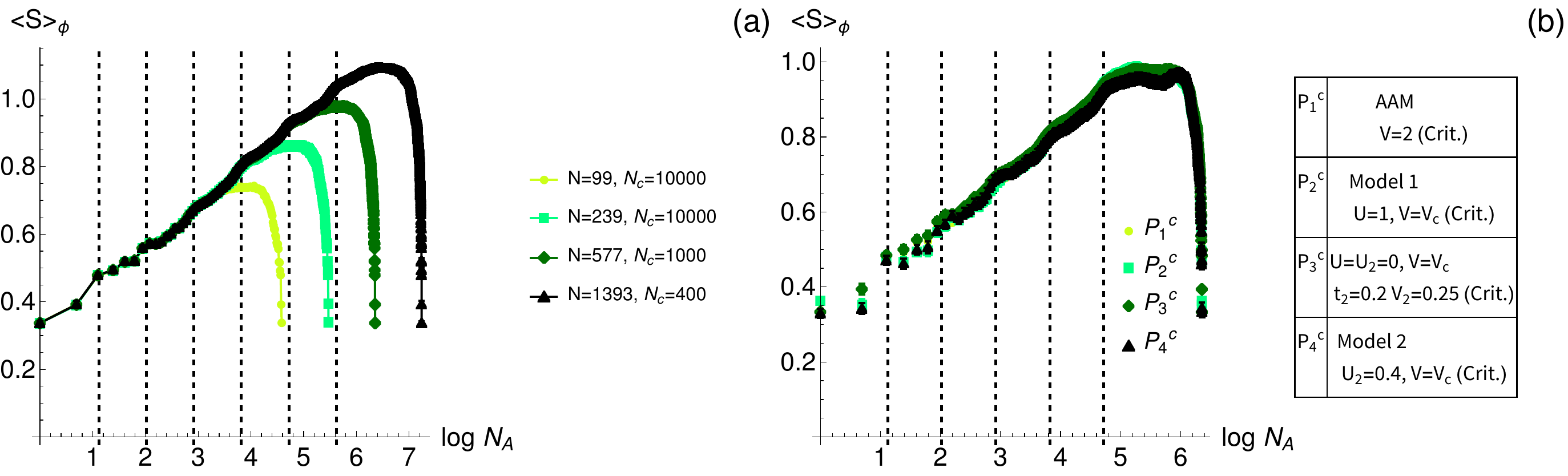}\caption{ (a) Entanglement entropy $\mathcal{S}$ as a function of the size of sub-system $A$, consisting of the first $N_A$ sites, for the non-interacting Aubry-André model. We used different system sizes and averaged over $N_c$ configurations of $\phi$ given by $\phi_j=2\pi j/N_c, \textrm{ },j=0,\cdots,N_c-1$. The vertical dashed lines are guides to the eye, showing the log-periodic maxima of the oscillations. (b) DMRG results at different critical points obtained for the parameter choices given in the table, averaging over $N_c \in [100-300]$ configurations. \label{fig:EE_log_periodic}}
\end{figure}

In the main text, we mentioned that the entanglement entropy, $\mathcal{S}$, shows showing log-periodic oscillations as a function of the subsystem size, at the critical point of the non-interacting Aubry-André model. In Fig.$\,$\ref{fig:EE_log_periodic}(a) we show the numerical results supporting this claim in a log-linear plot. By averaging $\mathcal{S}$ over a sufficiently large number of $\phi$-configurations, we see that these oscillations are robust to increasing the system size. In Fig.$\,$\ref{fig:EE_log_periodic}(b) we also show that these oscillations persist in the presence of interactions, at the critical point. 

\section{Duality transformation}

Here we build a many-body generalization of the duality transformation introduced in Ref.~\citep{HdualitiesScipost}. 
We start by writing the most occupied natural orbital as $\ket{\alpha=0}=\sum_{i}\psi_{i}^{(0)}\ket i$,
and defining its Fourier transform as

\begin{equation}
\tilde{\psi}_{k}^{(0),d}=\frac{1}{\sqrt{N}}\sum_{i=0}^{N-1}e^{\textrm{i}2\pi\tau_{c}ki}\psi_{i}^{(0)}.
\end{equation}

The hidden duality transformations defined in Ref.~\citep{HdualitiesScipost}
map points $(\phi,k)=(\phi_{0}+\Delta\phi,k_{0}+\Delta k)$ to points
$(\phi',k')=(\phi_{0}+\Delta k,k_{0}+\Delta\phi)$, where $(\phi_{0},k_{0})$
is the ``center'' of the hidden duality transformation. Setting
$(\phi_{0},k_{0})=(\varphi_{0},\kappa_{0})$, with $(\varphi_{0},\kappa_{0})$
given in the main text for the different used system sizes yields
a possible choice for which $\bm{\psi}^{(0)}\propto\tilde{\bm{\psi}}^{(0),d}$
at the self-dual point of the non-interacting Aubry-André model ($V=2$). For more generic
choices, we would need to compute $\psi_{i}^{(0)}$ at $(\phi,k)=(\phi_{0}+\Delta\phi,k_{0}+\Delta k)$
and $\tilde{\psi}_{k}^{(0),d}$ at $(\phi',k')=(\phi_{0}+\Delta k,k_{0}+\Delta\phi)$
to have $\bm{\psi}^{(0)}\propto\tilde{\bm{\psi}}^{(0),d}$.

\begin{figure}[h]
\centering{}\includegraphics[width=0.6\columnwidth]{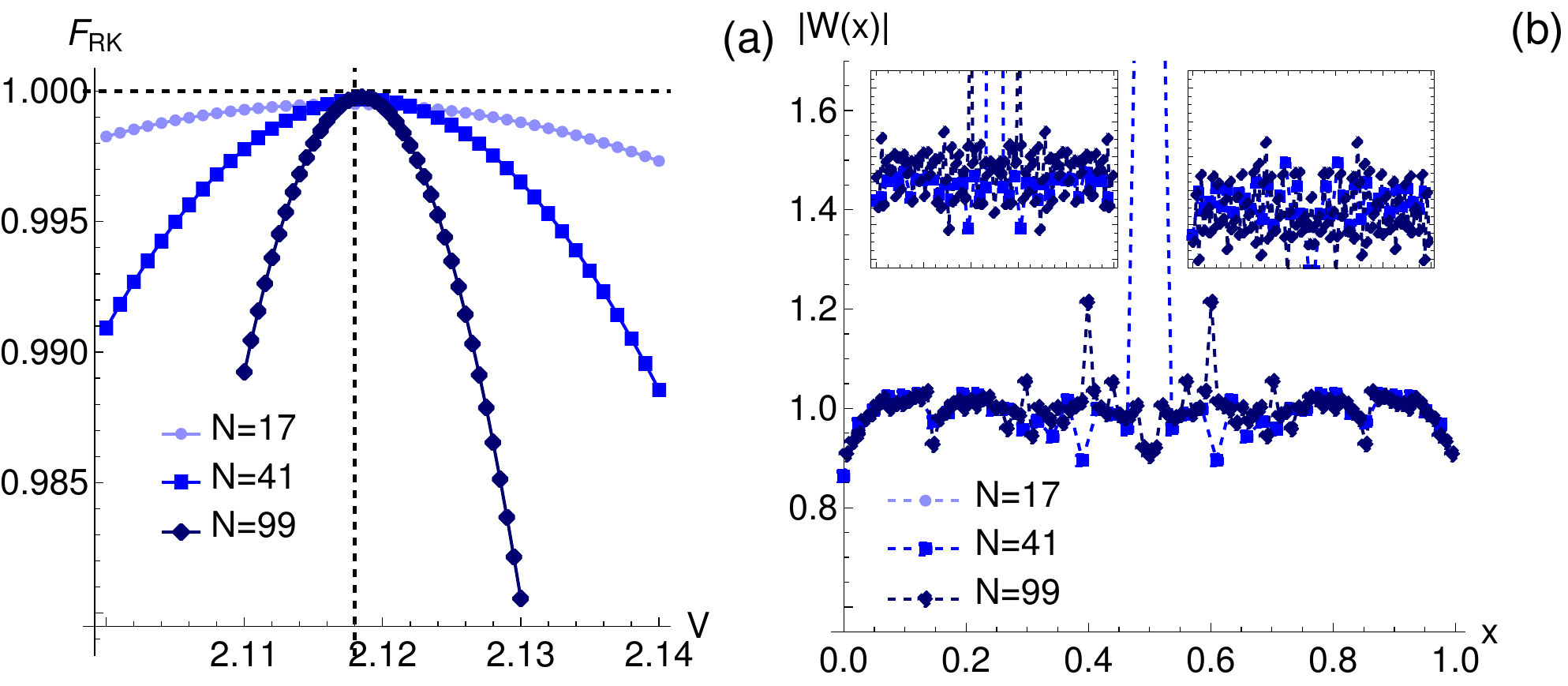}\caption{(a) $F_{RK}$ defined in the text for $U=0.5$ and $U_{2}=V_{2}=t_{2}=0$.
The dashed black line indicates $V=V_{c}(N=99)$. (b) Duality function
$\chi(x)$ introduced in Ref.$\,$\citep{10.21468/SciPostPhys.13.3.046}
defined through the natural orbital at $V=V_{c}(N)$. The insets show the results
if $\chi(x)$ was computed at the points marked in (a), slightly away
from the critical point: in this case we obtain a featureless function,
not robust to increasing $N$. 
\label{fig:duality_transformation}
}
\end{figure}

In Fig.$\,$\ref{fig:duality_transformation}(a)
we computed $F_{RK} \equiv (\tilde{\bm{\psi}}^{(0),d})^{*}\cdot\bm{\psi}^{(0)}$ using $(\phi,k)=(\phi_{0},k_{0})=(\varphi_{0},\kappa_{0})$ for model 1 with $U=0.5$ as an example. We see that $F_{RK}$ decreases with $N$, except when we cross
the critical point, where it becomes very close to $1$. This suggests
that $\bm{\psi}^{(0)}$ is almost equal to $\tilde{\bm{\psi}}^{(0),d}$
at this point. We can go one step further and define the duality transformation that relates $\bm{\psi}^{(0)}$ and $\tilde{\bm{\psi}}^{(0),d}$
at self-dual points as in Ref.$\,$\citep{10.21468/SciPostPhys.13.3.046}
(where the natural orbital replaces the role of the single-particle wave function).

From
$\bm{\psi}^{(0)}$ and $\bm{\tilde{\psi}}^{(0),d}$, we then define
the duality matrix $\mathcal{O}_{c}$ as in Ref.~\citep{HdualitiesScipost}:

\begin{equation}
\mathcal{O}_{c}[T^{n}\bm{\tilde{\psi}}^{(0),d}]=T^{n}\bm{\psi}_{i}^{(0)},\hspace{1em}n=0,\cdots,L-1,
\end{equation}
where $T$ is the cyclic translation operator defined as $T\psi=\psi'$
with $\psi'_{i}=\psi_{\mod(i+1,L)}$. Since $\mathcal{O}_{c}$ is
a circulant matrix, we may write it as

\begin{equation}
\mathcal{O}_{c}=U^{\dagger}WU
\end{equation}
where $U$ is a matrix with entries $U_{ij}=e^{2\pi{\rm i}\tau_{c}ij}$
and $W$ is a diagonal matrix $W_{ij}=w_{j}\delta_{ij}$ with the
eigenvalues $\{w_{j}\}$ of $\mathcal{O}_{c}$. We can therefore write

\begin{equation}
\bm{\psi}^{(0)}=U^{\dagger}W\bm{\psi}^{(0)}\leftrightarrow\psi_{i}^{(0)}=\sum_{\nu=0}^{L-1}e^{2\pi i\tau_{c}ij}w{}_{j}\psi_{j}^{(0)}.
\end{equation}

The eigenvalues $w_{j}$ are, as seen in Ref.~\citep{HdualitiesScipost},
evaluations of a function $W(x)$, that has period $\Delta x=1$,
at points $x_{j}=\mod\Big(j\tau_{c}+\frac{\phi}{2\pi},1\Big),\textrm{ }j=0,\cdots,L-1$.
This function is sampled in the whole interval $x\in[0,1[$ in the
limit that $\tau_{c}\rightarrow\tau$ ($N\rightarrow\infty$) and
encodes all the information on the duality transformation $W$. 
We show an example of the duality function $W(x)$ in Fig.$\,$\ref{fig:duality_transformation}(b),
where we see that a complicated function with features that are robust
to the increasing of $N$ is formed. 
 $W(x_{j})$ only has the meaning of a duality transformation
if $\bm{\psi}^{(0)}$ and $\tilde{\bm{\psi}}^{(0),d}$ are computed
at self-dual points (or at dual points in the extended and localized phases,
a case that was not considered here). We can however compute $W(x_{j})$
in the same way by using $\bm{\psi}^{(0)}$ and $\tilde{\bm{\psi}}^{(0),d}$
at any point, but in this case, since there is no duality transformation
connecting the wave functions, we expect $W(x_{j})$ to be featureless
and not robust for increasing system size. This is clearly shown in the insets
of Fig.$\,$\ref{fig:duality_transformation}(b).

\section{Generalized Chalker scaling and irrelevance of generic short-range interactions } 
We show that generic short-range (and some long-range) interactions are irrelevant at the critical point of the Aubry-André model in the $U\rightarrow 0$ limit, by unveiling the existence of a generalized Chalker scaling \citep{PhysRevLett.61.593,CHALKER1990253,PhysRevB.76.235119,PhysRevB.89.155140,PhysRevB.101.235121} at this point. All the results that we present in this section are for the parameters studied in the main text, namely $\tau= 1/\sqrt{2}$ and at half-filling, with $ N_{p}=\lfloor N/2\rfloor $ particles. Nonetheless, the technology here developed can be (and was) applied to more generic cases, as we comment at the end of the section. 

We consider the periodized form of the power-law interactions $U \sum_j j^{-w} n_{r}n_{r+j}$, given by 
\begin{equation}
    U\sum_{j=1}^{N-1}\sum_{k=0}^{\infty}(j+kN)^{-w}\sum_{r}n_{r}n_{r+j+kN} = U\sum_{j=1}^{N-1}N^{-w}\zeta(w,j/N)\sum_{r=1}^N n_{r}n_{r+j}, 
\end{equation}

\noindent  where $\zeta(w,y)= \sum_{k=0}^{\infty} (k+y)^{-w}$ is the Hurwitz zeta function and $ c_{r}^{\dagger}=c_{r+kN}^{\dagger},k\in \mathbb{Z}$ due to periodic boundary conditions. For such interaction, we can write the path integral for the grassman variables $\bar{c},c$ as 
\begin{equation}
Z=\int\mathcal{D}[\bar{c},c]e^{-(S_{0}[\bar{c},c]+S_{U}[\bar{c},c])}
\end{equation}
where, writing in the single-particle eigenbasis of the non-interacting Aubry-André model Hamiltonian $H_0$ (Eq.$\,$\ref{eq:H}, with $t_2=V_2=U=U_2=0$) with eigenenergies $\epsilon_{\alpha}=E_{\alpha}-\mu$ (measured relative to the chemical potential $\mu$),  we have

\begin{equation}
    S_{0}=\int_{0}^{\infty}d\tau\sum_{\alpha}\bar{c}_{\alpha}(\tau)(\partial_{\tau}+\epsilon_{\alpha})c_{\alpha}(\tau)
\end{equation}

\begin{equation}
S_{U}=-U\int_{0}^{\infty}d\tau\sum_{\alpha,\beta,\gamma,\delta}\bar{V}_{\alpha\beta\gamma\delta}\bar{c}_{\alpha}(\tau)\bar{c}_{\beta}(\tau)c_{\gamma}(\tau)c_{\delta}(\tau)
\end{equation}
\noindent and where $\bar{V}_{\alpha\beta\gamma\delta}=\Big(V_{\alpha\beta\gamma\delta}-V_{\beta\alpha\gamma\delta}+V_{\beta\alpha\delta\gamma}-V_{\alpha\beta\delta\gamma}\Big)/4$ is the antisymmetrized version of the interaction matrix elements

\begin{equation}
    V_{\alpha\beta\gamma\delta}=\sum_{r=1}^N \sum_{j=1}^{N-1}N^{-w}\zeta(w,j/N) \braket{\alpha|r}\braket{\beta|r+j}\braket {r|\gamma}\braket{r+j|\delta}
\end{equation}

\begin{figure}[h]
\centering{}\includegraphics[width=\columnwidth]{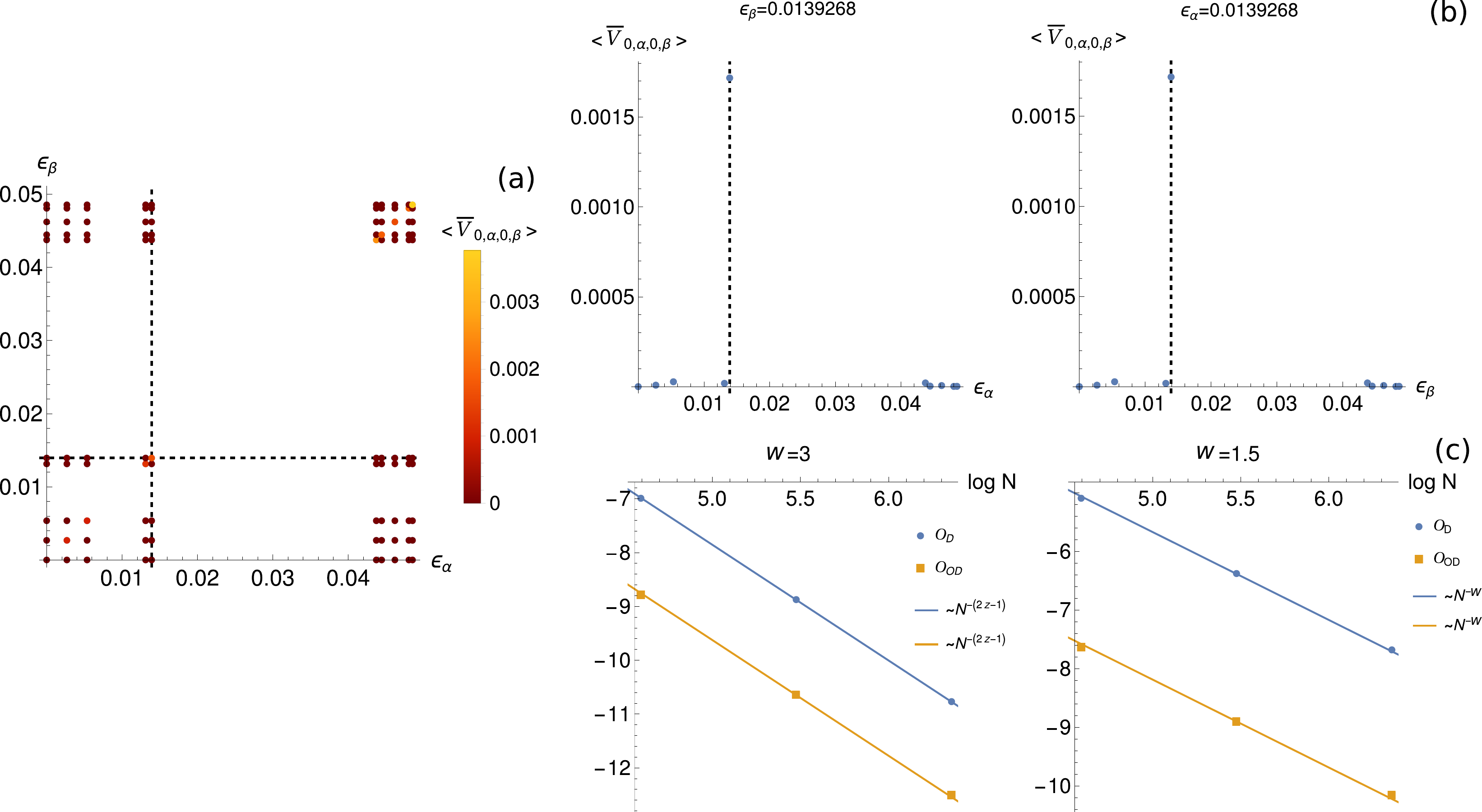}\caption{(a) $\langle \bar{V}_{0\alpha0\beta} \rangle_{\phi,\kappa}$ for $N=239$, $w=1.5$ and $\alpha,\beta=0,\cdots,9$, averaged over $200$ random configurations of $\phi$ and $\kappa$. We can see that the dominant contribution occurs for $\alpha=\beta$. This can also be seen in (b), where we make the cuts marked in (a) by the dashed lines. The vertical dashed line in (b) indicates the diagonal contribution, which is much larger than the remaining ones. In (c) we show that this conclusion is robust to increasing $N$. To do so, we compute the average diagonal, $ \mathcal{O}_{\textrm{OD}}=\frac{1}{n^{2}-n}\sum_{\alpha\neq\beta}^{n}\langle\bar{V}_{0\alpha0\beta}\rangle_{\phi,\kappa}$, and off-diagonal, $\mathcal{O}_{\textrm{D}}=\frac{1}{n}\sum_{\alpha=1}^{n}\langle\bar{V}_{0\alpha0\alpha}\rangle_{\phi,\kappa}$, contributions (fixing $n=9$, independently of $N$). Results are shown for $w=3$ (left) and $w=1.5$ (right) as examples.  $\mathcal{O}_{\textrm{OD}}$ and $\mathcal{O}_{\textrm{D}}$ scale identically with $N$, implying that the diagonal contribution dominates for any $N$.
\label{fig:ea_eb_diagonalDominance}
}
\end{figure}

We will now inspect the interacting part in detail. We have a 4-leg tensor on our hands. We want to study this tensor close to $\alpha,\beta,\gamma,\delta=0$, where 0 denotes the Fermi level. Since the tensor is antisymmetric, $\bar{V}_{0000}=0$. We can now inspect different combinations of indices to see how the 4-leg tensor behaves as the indices depart from 0. We can start by fixing 3 of the indices to be 0 and varying the remaining index. However, this yields zero due to antisymetry. We can also now fix 2 indices to 0 and vary the remaining 2 indices that we call $\alpha$ and $\beta$. The possible contributions are $\bar{V}_{0\alpha0\beta}$, $\bar{V}_{\alpha0\beta0}$=$\bar{V}_{0\alpha0\beta}$, $\bar{V}_{00\alpha\beta}=\bar{V}_{\alpha\beta00}=0$ and $\bar{V}_{\alpha00\beta}=\bar{V}_{0\alpha\beta0}=-\bar{V}_{0\alpha0\beta}$. Therefore, the only contribution that we need to compute is $\bar{V}_{0\alpha0\beta}$, as all the others are either zero or can be obtained from this one. In Fig.$\,$\ref{fig:ea_eb_diagonalDominance} we show that the most important contribution arises for $\alpha=\beta$ (we show examples for $w=1.5$ and $w=3$, but this remains true for other values of $w$). Therefore, we will focus on the contribution $\bar{V}_{0\alpha0\alpha}$. Note that higher-order contributions involve setting only one index to 0 and varying the others, but is already a contribution involving 3 energies, that we assume to be neglegible as $\alpha,\beta,\gamma,\delta\rightarrow0$. We then write the interacting part of the action as

\begin{equation}
S_{U}
=-4U\int_{0}^{\infty}d\tau\sum_{\alpha}\bar{V}_{0\alpha0\alpha}\bar{c}_{0}(\tau)\bar{c}_{\alpha}(\tau)c_{0}(\tau)c_{\alpha}(\tau)+\mathcal{O}(\epsilon_{\alpha}^{\mu}\epsilon_{\gamma}^{x})
\label{S_U}
\end{equation}

\noindent where we assumed that $\bar{V}_{0\alpha0\alpha}\sim\epsilon_{\alpha}^{\mu}$ and $\gamma$ denotes the additional index (or indices) that we choose to make finite in tensor $\bar{V}_{0\alpha0\alpha}$ [for instance $\bar{V}_{0\alpha\gamma\alpha}=\bar{V}_{0\alpha0\alpha}+\mathcal{O}(\epsilon_{\alpha}^{\mu}\epsilon_{\gamma}^{x})$] and the exponent $x$ may depend on this choice of indices. This contribution will therefore either be neglegible or the same as of $\bar{V}_{0\alpha0\alpha}$, if $x=0$. The term $\bar{V}_{0\alpha0\alpha}$ can be written explicitly as 

\begin{equation}
    \begin{array}{cc}
\bar{V}_{0\alpha0\alpha}= & \frac{1}{4}\sum_{j=1}^{N-1}N^{-w}\zeta(q,j/N)\Bigg(\sum_{r=1}^{N}(|\braket{0|r}|^{2}|\braket{\alpha|r+j}|^{2}+|\braket{\alpha|r}|^{2}|\braket{0|r+j}|^{2})\\
 & -\sum_{r=1}^N\braket{0|r}\braket{r|\alpha}\braket{\alpha|r+j}\braket{r+j|0}-\sum_{r=1}^N\braket{\alpha|r}\braket{r|0}\braket{0|r+j}\braket{r+j|\alpha}\Bigg)
\end{array}
\end{equation}

\begin{figure}[h]
\centering{}\includegraphics[width=\columnwidth]{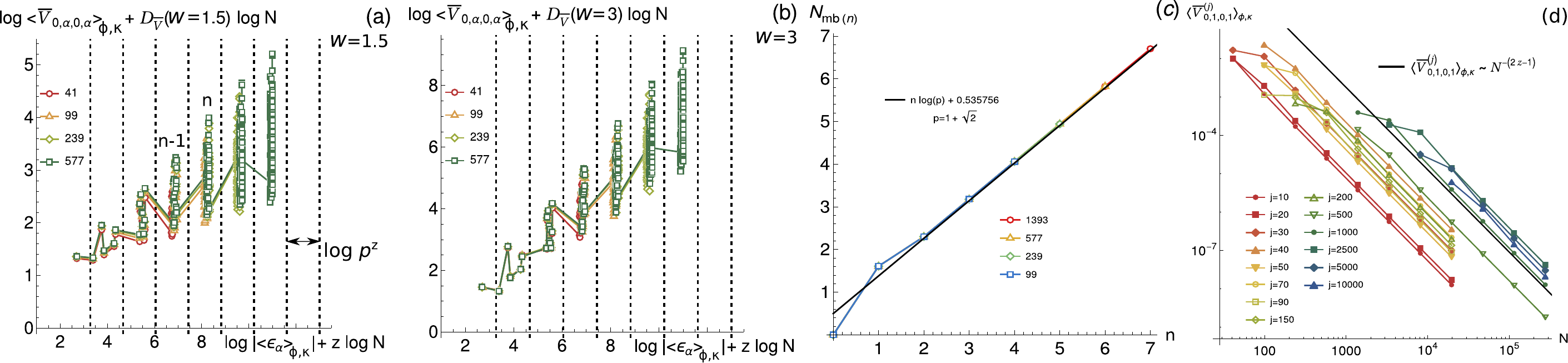}\caption{
(a,b) $\langle \bar{V}_{0\alpha0\alpha} \rangle_{\phi,\kappa}$ averaged over 200 random configurations of $\phi$ and $\kappa$ ($\langle\rangle_{\phi,\kappa}$ denotes the average over $\phi,\kappa$ configurations) for $w=1.5$ (a) and $w=3$ (b), where $D_{\bar{V}}(w=1.5)=w$ and $D_{\bar{V}}(w=3)=2z-1$, with $z=1.575$ being the dynamical critical exponent. (c)  Number of states in each miniband, $N_{{\rm mb}(n)}$. Note that the minibands are only well-defined (that is, there are clear clusters of states) for $n \geq 2$, above which the  scaling $N_{{\rm mb}(n)} \propto p^n$ is observed. The dashed lines in (a,b) are separated by $\log p^z$, implying that the average energy of each miniband scales as $p^{zn}$. (d) $\langle \bar{V}^{(j)}_{0\alpha0\alpha} \rangle_{\phi,\kappa}$ for the interaction term in Eq\,\ref{eq:j_interaction}, and for $\alpha=1$, as a function of system size $N$ and averaged over 4800 random configurations of $\phi$ and $\kappa$. 
\label{fig:collapses_SM}
}
\end{figure}

In Figs.\,\ref{fig:collapses_SM}(a,b), we show that it is possible to collapse the results for $\bar{V}_{0\alpha0\alpha}$ for different approximant system sizes and different energies. The collapse becomes better as $\epsilon_{\alpha}\rightarrow0$. Furthermore, there are clusters of eigenvalues that form on the $\log|\epsilon_{\alpha}|$ scale, that we will can “minibands” in the following. In Fig.\,\ref{fig:collapses_SM}(c) we can see that the number of states in each miniband scales as $N_{{\rm mb}(n)}\propto p^{n}$. By realizing that increasing the order of system size approximant introduces a new miniband, we can easily find that $p=N_{m+1}/N_m \rightarrow1+\sqrt{2}$ as $m\rightarrow\infty$ where $N_{m}$ is the m-th order system size approximant for $\tau=1/\sqrt{2}$. By defining $\bar{\epsilon}_{n}^{(m)}=N_{{\rm mb}(n)}^{-1}\sum_{\alpha\in{\rm mb(n)}}\epsilon_{\alpha}^{(m)}$ (where the superscript “(m)” indicates the eigenenergies for the m-th order size approximant), we also have that $\bar{\epsilon}_{n}\propto p^{nz}$, as indicated in Fig.\,\ref{fig:collapses_SM}(a), where $z=1.575$ is the dynamical critical exponent. Naturally, the scaling collapse in this figure also implies that $\epsilon_{\alpha}^{(m-l)}=p^{zl}\epsilon_{\alpha}^{(m)}$. These observations allow us to write the following ansatz,

\begin{equation}
    \bar{V}_{0\alpha0\alpha}=\mathcal{C}N_{m}^{-D_{\bar{V}}}(|\epsilon_{\alpha}^{(m)}|N_{m}^{z})^{\mu}\sum_{n}f([|\epsilon_{\alpha}^{(m)}|-\bar{\epsilon}_{n}^{(m)}]N_{m}^{z})
    \label{V0a0a_ansatz}
\end{equation}

\noindent where $\mathcal{C}$ is some constant independent of energy and $N_{m}$. We note that, as shown in Figs.\,\ref{fig:collapses_SM}(a,b) and in the main text, $D_{\bar{V}}$ depends on $w$. We will discuss this dependence below in more detail below. At this point we also note that when averaged over minibands, $ \bar{V}_{0\alpha0\alpha} \sim (|\epsilon_{\alpha}^{(m)}|)^{\mu}$, where $\mu > 0$. This shows that there is a generalized Chalker scaling \citep{PhysRevLett.61.593,CHALKER1990253,PhysRevB.76.235119,PhysRevB.89.155140,PhysRevB.101.235121} at the critical point of the Aubry-André model, manifested by power-law correlations (on average) between the single-particle eigenfunctions with respect to their energy difference. 

To carry out a power-counting analysis and inspect the scaling dimension of the interactions, we take a large enough system size to begin with so that the data collapse is quite good for the relevant energies of choice and Eq.\,\ref{V0a0a_ansatz} holds. In each renormalization-group (RG) step, we throw away a miniband and rescale the energies. Starting with an energy cutoff $\Lambda_{k}$, after $l$ RG steps we end up with a cutoff $\Lambda_{k+l}=\Lambda_{k}/p^{zl}$. We also start with an initial system size $N_{m}$. The non-interacting action $S_{0}$, after introducing the cutoff, is given by 

\begin{equation}
    S_{0}^{\Lambda_{k},N_{m}}=\int_{0}^{\infty}d\tau\int_{-\Lambda_{k}}^{\Lambda_{k}}d\epsilon\textrm{ }\sum_{\alpha}\delta(\epsilon-\epsilon_{\alpha}^{(m)})\bar{c}(\epsilon,\tau)(\partial_{\tau}+\epsilon)c(\epsilon,\tau)
\end{equation}

 After after $l$ RG steps, it becomes:

\begin{equation}
S_{0}^{\Lambda_{k+l},N_{m}}=\int_{0}^{\infty}d\tau'\int_{-\Lambda_{k}}^{\Lambda_{k}}d\epsilon'\textrm{ }\sum_{\alpha}\delta(\epsilon'-\epsilon_{\alpha}^{(m-l)})\bar{c}(\epsilon',\tau')\Big(\partial_{\tau'}+\epsilon'\Big)c(\epsilon',\tau')=S_{0}^{\Lambda_{k},N_{m-l}}    
\end{equation}

\noindent where we used $\ensuremath{\epsilon'=p^{zl}\epsilon}\textrm{ , \ensuremath{\tau'=\tau p^{-zl}}}, \epsilon_{\alpha}^{(m-l)}=p^{zl}\epsilon_{\alpha}^{(m)}$ and defined $\bar{c}(\epsilon'p^{-zl},\tau'p^{zl})=\bar{c}(\epsilon',\tau');\textrm{ }c(\epsilon'p^{-zl},\tau'p^{zl})=c(\epsilon',\tau')$. The new action after $l$ RG steps therefore corresponds to the same action, but for a smaller system size $N_{m-l}$. For the interacting part, we have 

\begin{equation}
S_{U}^{\Lambda_{k},N_{m}}=-U\mathcal{C}N_{m}^{-D_{\bar{V}}}\sum_{\alpha}\int_{-\Lambda_{k}}^{\Lambda_{k}}d\epsilon\delta(\epsilon-\epsilon_{\alpha}^{(m)})(|\epsilon|N_{m}^{z})^{\mu}\sum_{n}f([|\epsilon|-\bar{\epsilon}_{n}^{(m)}]N_{m}^{z})\bar{c}(0,\tau)\bar{c}(\epsilon,\tau)c(0,\tau)c(\epsilon,\tau)
\end{equation}

\noindent where the factor 4 in Eq\,\ref{S_U} was absorved in the constant $\mathcal{C}$. The full action for the interacting part after $l$ RG steps is therefore 

\begin{equation}
    \begin{array}{cc}
S_{U}^{\Lambda_{k+l},N_{m}}=-U\mathcal{C}p^{-(D_{\bar{V}}-z)l}N_{m-l}^{-(D_{\bar{V}}-z)}\int d\tau'\sum_{\alpha}\int_{-\Lambda_{k}}^{\Lambda_{k}}d\epsilon'\textrm{ }\delta(\epsilon'-\epsilon_{\alpha}^{(m-l)})(|\epsilon'|N_{m-l}^{z})^{\mu}\\
\times\sum_{n}f([|\epsilon'|-\bar{\epsilon}_{n}^{(m-l)}]N_{m-l}^{z})\bar{c}(0,\tau')\bar{c}(\epsilon',\tau')c(0,\tau')c(\epsilon',\tau')\\
=p^{-(D_{\bar{V}}-z)l}S_{U}^{\Lambda_{k},N_{m-l}}
\end{array}
\end{equation}

In summary, after $l$ RG steps we have:

\begin{equation}
S^{\Lambda_{k+l},N_{m}}=S_{0}^{\Lambda_{k+l},N_{m}}+S_{U}^{\Lambda_{k+l},N_{m}}=S_{0}^{\Lambda_{k},N_{m-l}}+p^{-(D_{\bar{V}}-z)l}S_{U}^{\Lambda_{k},N_{m-l}}
\end{equation}

This implies that interactions the scaling dimension of the interacting part is $z-D_{\bar{V}}$, and therefore interactions are irrelevant when $D_U = D_{\bar{V}}>z$. In Fig\,\ref{fig:4}(b) of the main text, we have seen that the thermodynamic-limit behaviour of $D_{\bar{V}}(w)$ is compatible with

\begin{equation}
D_{\bar{V}}(w)=\begin{cases}
w & ,w<2z-1\\
2z-1 & ,w\geq2z-1
\end{cases}
\end{equation}

This implies that interactions are irrelevant for $w>z$, marginal for $w=z$ and relevant for $w<z$. The relevance of interactions for $w<z$ is left for future exploration. These results also imply that even when long-range interactions are considered, they can be irrelevant in the $U \rightarrow 0$ limit if they decay fast enough. On the other hand, it also follows that short-range interactions have $D_{\bar{V}}=2z-1$ and therefore their scaling dimension is $D_U=1-z$. Since $z>1$ at the critical point, short-range interactions are irrelevant.  At the extended phase, on the other hand, $z=1$, which implies that interactions are marginal, in agreement with the $V=0$ results. 

To show that short-range interactions are irrelevant in more detail, we consider the following finite-range interacting terms (again assuming periodic boundary conditions),

\begin{equation}
    H_U^{(j)}= U \sum_{r=1}^N n_r n_{r+j}
    \label{eq:j_interaction}
\end{equation}

\noindent and compute the associated antisymmetrized interaction $\bar{V}^{(j)}_{0 1 0 1} \propto N^{-D_{\bar{V}}}$ for each interaction term of this type, in Fig.\,\ref{fig:collapses_SM}(d). We find that no matter the interaction range $j$, if the system size becomes sufficiently larger than this range $D_{\bar{V}}(N) \rightarrow 2z-1$. Therefore, any short-range function of these interaction terms should also follow this behaviour. With this in mind, we expect that the universal behaviour unveiled in this work is not restricted to the interactions studied in Eq\,\ref{eq:H}, but also holds for more generic short-range (and even some long-range) interactions. Even though in this section we focused on the choices of parameters used in the main text, we checked that the same conclusions can also be drawn for other fillings and other values of $\tau$ also considered in \citep{PhysRevB.101.174203}. 

We finish this section by showing that the short-range dimension $D_{\bar{V}}=2z-1$ can be understood from simple arguments. We start by writing 

\begin{equation}
    \bar{V}_{0101}=N^{-D_{\bar{V}}}f(\epsilon_{g}N^{z})
\end{equation}

\noindent assuming that $D_{\bar{V}}$ is unknown, where $\epsilon_{g}$ is the energy gap for a system size $N$. Since we have $\epsilon_{g}=KN^{-z}$, where $K$ is a constant, we know that $\bar{V}_{0101}=N^{-D_{\bar{V}}}f(K)\sim N^{-D_{\bar{V}}}$ and therefore 

\begin{equation}
    \bar{V}_{0101}=\epsilon_{g}^{D_{\bar{V}}/z}
    \label{eq:eg1}
\end{equation}

On the other hand, we can write

\begin{equation}
\bar{V}_{0101}=  \sum_{r=1}^N \bar{V}_{\epsilon_{g}}^{r}\end{equation}

\noindent where

\begin{equation}
\begin{aligned}
\bar{V}_{\epsilon_{g}}^{r}=\frac{1}{4}\Bigg(|\braket{ 0|r}|^{2} |\braket{1|r+j}|^{2}+|\braket{1|r}|^{2}|\braket{ 0|r+j}|^{2}\\
-(\braket{0|r} \braket{r|1}  \braket{1|r+j} \braket{r+j|0} +\textrm{c.c})\Bigg)
\end{aligned}
\end{equation}

After averaging over $\phi$ and $\kappa$, translational invariance is restored and $\bar{V}_{\epsilon}^{r}$ becomes r-independent. Furthermore, we know that $\bar{V}_{\epsilon_{g}}^{r}\rightarrow0$ as $\epsilon_{g}\rightarrow0$. Expanding $\bar{V}_{\epsilon_{g}}^{r}$ in powers of $\epsilon_{g}$, assuming it to be a regular function:

\begin{equation}
\bar{V}_{\epsilon_{g}}^{r}=a_{1}\epsilon_{g}+a_{2}\epsilon_{g}^{2}+\cdots
\end{equation}

We have that $a_{1}=0$ since it can be easily shown that $\bar{V}_{\epsilon_{g}}^{r}\geq0$ for any $\epsilon_{g}$. We therefore have 

\begin{equation}
\bar{V}_{0101}\sim  N\epsilon_{g}^{2}\sim\epsilon_{g}^{2-1/z}
\end{equation}

By comparing with Eq.\,\ref{eq:eg1}, this therefore implies that  $D_{\bar{V}}=2z-1$. Therefore, we conclude that the scaling dimension for short-range interactions simply follows from $\bar{V}_{\epsilon_{g}}^{r}$ being a regular function of $\epsilon_{g}$.

\section{Charge gap scaling for alternative choices of $\tau$ and filling $\rho$}

From the results that we obtained in the main text, we have seen that
the scalings of the charge gap (and other quantities such as the fractal
dimension) with system size obtained at different LL-AG transitions are compatible,
no matter the chosen parameters (hoppings, potential, interactions),
at half-filling ($\rho=1/2$) and for approximants of $\tau=1/\sqrt{2}$
{[}Fig.$\,$3 of the main text{]}. A natural question that arises
is whether this is a special feature of our choice of $\rho$ and
$\tau$. In particular, we know that the dynamical exponent $z$ depends
on both $\rho$ and $\tau$ in the non-interacting limit, for the
Aubry-André model \cite{PhysRevB.101.174203}. If we make other choices of $\rho$
and $\tau$, is the charge gap scaling also independent on the remaining
Hamiltonian parameters, as long as we are at the critical point? Since
this is a question that we can already ask in the non-interacting
limit, we will take the class of models considered in the main text, in the non-interacting limit, with Hamiltonian given by:

\begin{equation}
\begin{aligned}H= & -\sum_{i}c_{i}^{\dagger}c_{i+1}+t_{2}\sum_{i}c_{i}^{\dagger}c_{i+2}+{\rm h.c.}\\
 & +\sum_{i}\Big(V\cos(2\pi\tau_{c}i+\phi)+V_{2}\cos[2(2\pi\tau_{c}i+\phi)]\Big)c_{i}^{\dagger}c_{i}
\end{aligned}
\textrm{ GAA model}
\end{equation}

\begin{figure}
\centering{}\includegraphics[width=1\columnwidth]{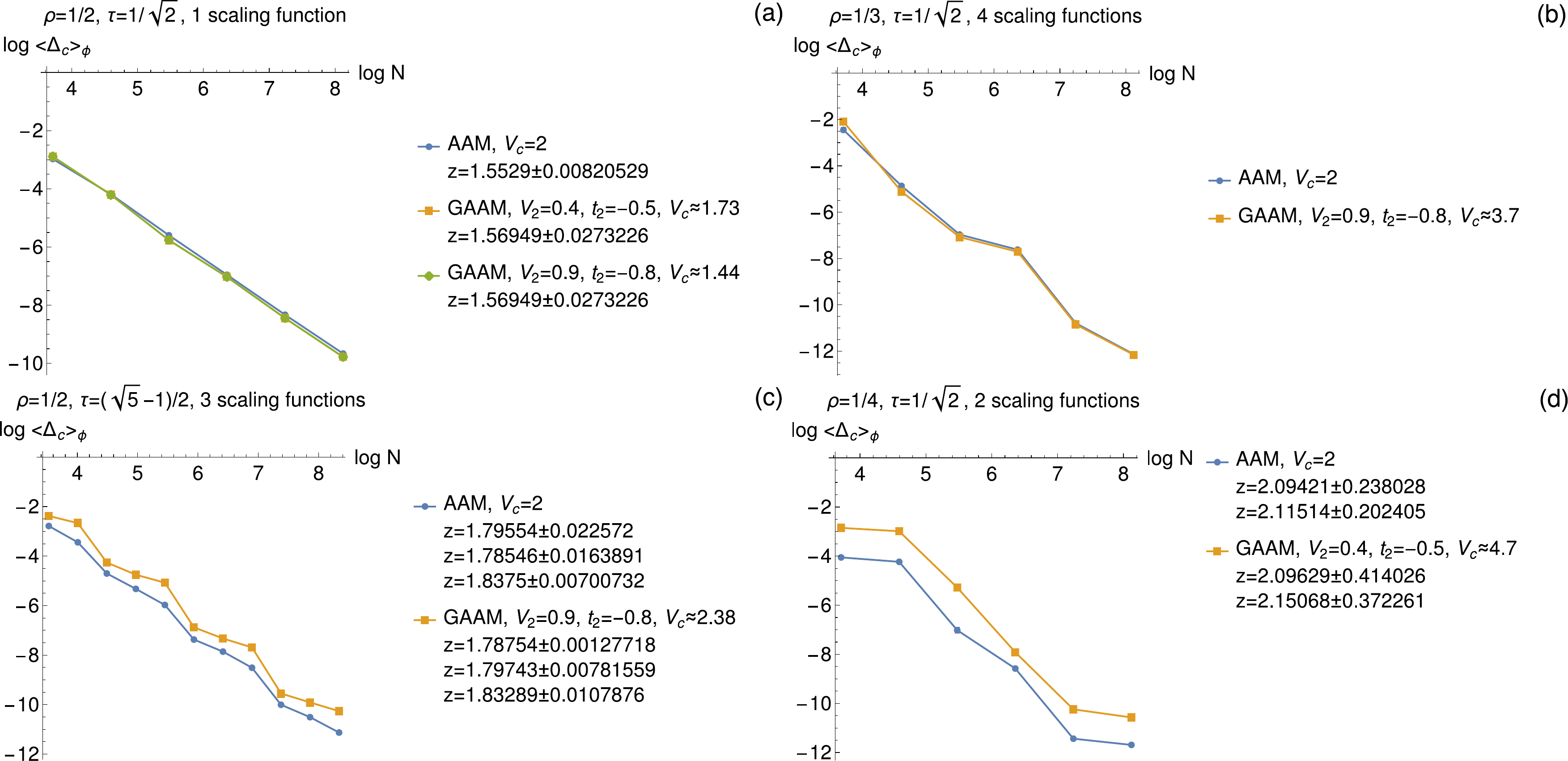}\caption{Scaling of the charge gap using open boundary conditions and averaged
over $\phi$, for different fillings and choices of $\tau$ and at
critical points of different models. The different critical points
were estimated by imposing twisted boundary conditions and estimating
the value $V=V_{c}$ for which $\Delta E_{\kappa\varphi}=1$ (leaving
the remaining parameters fixed), for the largest used system size.
We indicate all the relevant parameters in the figure and also the
number of scaling functions for the selected $\tau$ and $\rho$,
that were obtained in \cite{PhysRevB.101.174203}. The values of $z$
indicated are extracted from fits using sizes $N_{\mathcal{N}j+1},j=0,\cdots,j_{{\rm max}}$
of the lists given in table \ref{tab:size_choice_z_non_interacting},
where $\mathcal{N}$ is the number of scaling functions. In (a) we
used the $5$ largest sizes, in (c) the 3 largest sizes belonging
to each scaling function and in (d) we used all sizes. The different
$z$ estimates in (c,d) for the same critical points were obtained
from fits to sizes that belong to the different existing scaling functions.
\label{fig:cgap_scaling_non_interacting_ManyModels}}
\end{figure}

For the finite-size scaling results that follow, we use open boundary
conditions and the sizes and rational approximants $\tau_{c}$ given
in table \ref{tab:size_choice_z_non_interacting}. The results are
given in Fig.$\,$\ref{fig:cgap_scaling_non_interacting_ManyModels},
where we can see that the scalings obtained at critical points of
widely different models are very compatible for fixed $\rho$ and
$\tau$. In some cases, there are more than one scaling functions, which
means that an accurate finite-size scaling analysis should consider
the system sizes that belong to the different scaling functions separately
\cite{PhysRevB.101.174203}. Remarkably, even the scaling features
that arise due to the existence of multiple scaling functions (e.g.,
the 3-step scaling in Fig.$\,$\ref{fig:cgap_scaling_non_interacting_ManyModels}
due to the existence of 3 scaling functions) holds at different critical
points as long as $\rho$ and $\tau$ are fixed. These results support
our claim that the scaling invariance that we observed at the critical
point is not a special feature of our choice of $\rho$ and $\tau$. We checked for additional models, e.g. the model in Ref.$\,$\cite{PhysRevLett.114.146601}, and obtained compatible results.
\begin{center}
\begin{table}[h]
\begin{centering}
\begin{tabular}{|c|c|c|c|c|c|c|}
\hline 
\multicolumn{7}{|c|}{$\tau=1/\sqrt{2}$}\tabularnewline
\hline 
$N$ & 41 & 99 & 239 & 577 & 1393 & 3363\tabularnewline
\hline 
\hline 
$\tau_{c}$ & $\frac{29}{41}$ & $\frac{70}{99}$ & $\frac{169}{239}$ & $\frac{408}{577}$ & $\frac{985}{1393}$ & $\frac{2378}{3363}$\tabularnewline
\hline 
$N_{c}$ & 500 & 500 & 500 & 500 & 500 & 300\tabularnewline
\hline 
\end{tabular}
\vspace{0.2cm}
\par\end{centering}
\begin{centering}
\begin{tabular}{|c|c|c|c|c|c|c|c|c|c|c|c|}
\hline 
\multicolumn{12}{|c|}{$\tau=(\sqrt{5}-1)/2$}\tabularnewline
\hline 
$N$ & 34 & 55 & 89 & 144 & 233 & 377 & 610 & 987 & 1597 & 2584 & 4181\tabularnewline
\hline 
\hline 
$\tau_{c}$ & $\frac{21}{34}$ & $\frac{34}{55}$ & $\frac{55}{89}$ & $\frac{89}{144}$ & $\frac{144}{233}$ & $\frac{233}{377}$ & $\frac{377}{610}$ & $\frac{610}{987}$ & $\frac{987}{1597}$ & $\frac{1597}{2584}$ & $\frac{2584}{4181}$\tabularnewline
\hline 
$N_{c}$ & 1000 & 1000 & 1000 & 1000 & 750 & 750 & 500 & 500 & 500 & 300 & 250\tabularnewline
\hline 
\end{tabular}
\par\end{centering}
\caption{Choices of sizes $N$, rational approximants $\tau_{c}$ and number
of $\phi$-configurations $N_{c}$ for $\tau=1/\sqrt{2}$ and $\tau=(\sqrt{5}-1)/2$.
The different phases $\phi$ were chosen from a uniform grid given
by $\phi_{j}=2\pi j/N_{c},j=0,\cdots,N_{c}-1$. \label{tab:size_choice_z_non_interacting}}
\end{table}
\par\end{center}

\end{document}